\documentclass[graybox]{svmult}

\usepackage{type1cm}        % activate if the above 3 fonts are
\usepackage{makeidx}         % allows index generation
\usepackage{graphicx}        % standard LaTeX graphics tool
\usepackage{multicol}        % used for the two-column index
\usepackage[bottom]{footmisc}% places footnotes at page bottom

\usepackage{newtxtext}       % 
\usepackage[varvw]{newtxmath}       % selects Times Roman as basic font
\usepackage{url}

\makeindex             % used for the subject index
\DeclareMathOperator*{\esssup}{ess\,sup}

\begin{document}

\title*{Quadratic forms for Aharonov-Bohm Hamiltonians}
\author{Davide Fermi}
\institute{Davide Fermi \at
Universit\`a degli Studi Roma Tre, Dipartimento di Matematica e Fisica, L.go S. L. Murialdo 1, 00146 Roma, Italy, \email{davide.fermi@uniroma3.it}, webpage: \url{https://fermidavide.com}}

\maketitle

\vspace{-0.5cm}
\abstract{We consider a charged quantum particle immersed in an axial magnetic field, comprising a local Aharonov-Bohm singularity and a regular perturbation. Quadratic form techniques are used to characterize different self-adjoint realizations of the reduced two-dimensional Schr\"odinger operator, including the Friedrichs Hamiltonian and a family of singular perturbations indexed by $2 \!\times\! 2$ Hermitian matrices. The limit of the Friedrichs Hamiltonian when the Aharonov-Bohm flux parameter goes to zero is discussed in terms of $\Gamma$\! - convergence.}
\vspace{0.2cm}



\section{Introduction}

In a pioneering work dating back to 1949 \cite{ES49}, Ehrenberg and Siday foretold that charged quantum particles confined within regions where the electromagnetic field vanishes do still experience a phase shift, which can be expressed in terms of non-zero electromagnetic potentials. Their prediction did not attract great attention until Aharonov and Bohm re-discovered it in an independent research of 1959 \cite{AB59}, eventually reaching a much larger audience. Ever since then, people referred to the said phenomenon as the ``Aharonov-Bohm effect'' \cite{BT09,PT89}.
Despite sound experimental evidence \cite{TOMKEYY86}, this groundbreaking discovery generated conflicting views and a long-standing debate about the reality of electromagnetic potentials and the tenability of the locality principle in quantum mechanics \cite{AB61}. This controversy somehow continues even nowadays \cite{ACR16,K15,V12}, though there are strong indications that an explanation in terms of local field interactions can actually be attained within the framework of QED \cite{K22,MV20}.

The prototypical Aharonov-Bohm configuration consist of a single, non-relativistic, spinless and electrically charged quantum particle moving outside of a long thin solenoid. More precisely, attention is restricted to a low-energy regime in which the De Broglie wavelength of the particle is much larger than the solenoid section diameter and, at the same time, much smaller than the solenoid longitudinal length. The natural first order approximation considers an ideal solenoid of infinite length, zero cross section and finite total magnetic flux. Against this background, the Schr\"odinger operator ruling the dynamics of the particle reads
	\begin{equation*}
		H_{\mbox{{\scriptsize 3D}}} = {1 \over 2m} \big(-i \hbar \nabla - q \mathbf{A}_{\mbox{{\scriptsize 3D}}}\big)^2\,,
		\quad\;\;
		\mathbf{A}_{\mbox{{\scriptsize 3D}}}(x,y,z) = {\Phi \over 2\pi}\left(-\,{y \over x^2+y^2}\,,\, {x \over x^2+y^2}\,,0\right),	
	\end{equation*}
where $\hbar$ is the reduced Plank constant, $m$ is the particle mass, $q$ is the electric charge, $\Phi$ is the total magnetic flux across the solenoid, and $(x,y,z)$ are coordinates in $\mathbb{R}^3$ such that the solenoid coincides with the $z$-axis. The corresponding singular magnetic field is given by $\mathbf{B} = \mbox{curl}\, \mathbf{A}_{\mbox{{\scriptsize 3D}}} = (0,0,\Phi\,\delta_{(x,y) \,=\, (0,0)})$.

Upon factorizing the axial direction and passing to natural units of measure, the model is described by the reduced Schr\"odinger operator
	\begin{equation}\label{eq:Ha0}
		H_{\alpha} := \big(-i \nabla + \mathbf{A}_{\alpha}\big)^2\,, \qquad \mathbf{A}_{\alpha}(\mathbf{x}) := \alpha\,{\mathbf{x}^{\perp} \over |\mathbf{x}|^2}\,,
	\end{equation}
acting in $L^2(\mathbb{R}^2)$. Here and in the sequel, $\mathbf{x} = (x,y) \in \mathbb{R}^2$ and $\mathbf{x}^{\perp} \equiv (-y,x)$. Besides, $\alpha$ is a dimensionless parameter measuring the magnetic flux $\Phi$ in units of the flux quantum $2\pi\hbar/q$. It entails no loss of generality to assume\footnote{For any $\alpha \!\in\! \mathbb{R}$, consider the decomposition $\alpha = 2\ell + \tilde{\alpha}$ with $\ell \!\in\! \mathbb{Z}$, $\tilde{\alpha} \!\in\! (-1,1)$. For any fixed determination of $\arctan$, the map $(U \psi)(\mathbf{x}) \!=\! e^{-2i \ell \arctan(x/y)} \psi(\mathbf{x})$ defines a unitary operator in $L^2(\mathbb{R}^2)$. A direct computation gives $ U \big(- i \nabla + \alpha {\mathbf{x}^{\perp} \over |\mathbf{x}|^2}\big)^2 U^{-1} \!=\! \big(- i \nabla + (\alpha - 2 \ell) {\mathbf{x}^{\perp} \over |\mathbf{x}|^2}\big)^2$, showing that $H_{\alpha}$ is unitarily equivalent to $H_{\tilde{\alpha}}$ with $\tilde{\alpha} \!\in\! (-1,1)$. The condition $\tilde{\alpha} \!\in\! [0,1)$ can then be realized exploiting conjugation symmetry. The case $\alpha \!=\! 0$ is here discarded because of its triviality ($H_{\alpha \,=\, 0}$ is just the Laplacian in $\mathbb{R}^2$, with no magnetic flux).}
	\begin{equation*}
		\alpha \in (0,1)\,.	
	\end{equation*}

Due to the singularity of $\mathbf{A}_{\alpha}$ at $\mathbf{x}\! =\! \mathbf{0}$, the self-adjointness of $H_{\alpha}$ is not granted a priori. Assessing this feature is in fact a crucial task.
Decomposing in angular harmonics and exploiting the exact solvability of the radial problems, all admissible self-adjoint extensions of the symmetric restriction $H_{\alpha} \!\!\upharpoonright\! C^{\infty}_{c}(\mathbb{R}^2 \setminus \{\mathbf{0}\})$ were originally characterized via Krein-von Neumann methods in \cite{AT98} and \cite{DS98} (see also \cite{BDG11,DFNR20,PR11}). These extensions include the Friedrichs one and finite rank perturbations thereof, forming a family labeled by $2 \!\times\! 2$ complex Hermitian matrices.
It is worth noting that a complete analysis of the spectral and scattering properties of the resulting Hamiltonians can be derived by resolvent techniques.\linebreak
A complementary approach relies on considering first finite size solenoids, partially shielded by electrostatic potentials, and then examining suitable scaling regimes \cite{OP08,Kr64,MVG95,Ta99,Ta01}. Different self-adjoint realizations of $H_{\alpha}$ are obtained as limits (in strong resolvent sense) of Hamiltonian operators comprising just regular potentials. In this connection, zero-energy resonances produced by the shielding electrostatic potential play a key role. This also allows to gain some intuition about the physical meaning of different self-adjoint realizations.

The present work examines magnetic perturbations of the pure Aharonov-Bohm Hamiltonian $H_{\alpha}$, continuing the analysis begun in \cite{CO18,CF20}. The main goal is to characterize self-adjoint realizations in $L^2(\mathbb{R}^2)$  of the Schr\"odinger operator
	\begin{equation*}
		H_{\alpha,S} := (-i \nabla + \mathbf{A}_{\alpha} + \mathbf{S})^2\,,
	\end{equation*}
where $\mathbf{A}_{\alpha}$ is like in \eqref{eq:Ha0} and $\mathbf{S} \!\in\! L^{\infty}_{\mbox{{\scriptsize loc}}}(\mathbb{R}^2,\mathbb{R}^2)$ is the vector potential associated to a regular, external axial magnetic field, to be regarded as a perturbation of the Aharonov-Bohm singular flux. The said perturbation allows to account, in particular, for magnetic traps and leakages of magnetic field lines outside of the solenoid coils.

The case of a homogeneous magnetic field, corresponding to $\mathbf{S} = {B \over 2}\,\mathbf{x}^{\perp}$ ($B \!>\! 0$\linebreak being the magnetic field intensity), was previously analyzed in \cite{ESV02}. An exhaustive classification of all self-adjoint realizations of $H_{\alpha,S}$ (with the said choice of $\mathbf{S}$) and of their spectral properties was derived therein by means of Krein-von Neumann methods, exploiting again decomposition in angular harmonics.

For a generic perturbation $\mathbf{S}$, lacking rotational symmetry, the decomposition in angular harmonics appears to be somewhat artificial and the Krein-von Neumann construction cannot be implemented straightaway. On the contrary, a quadratic form approach is more natural and flexible. In the context under analysis the use of quadratic forms was first proposed in \cite{CO18} for the pure Aharonov-Bohm setting\linebreak with $\mathbf{S} \!=\! \mathbf{0}$. In \cite{CF20} similar techniques were employed to characterize the Friedrichs realization of $H_{\alpha,S}$ and a one-parameter family of singular $s$-wave perturbations. 

In this paper we extend the previous analysis, including $p$-wave and mixed singular perturbations of the Friedrichs Hamiltonian (see Theorem \ref{thm:QasB} and Corollary \ref{cor:HasB}). In view of the results derived in \cite{AT98,DS98,ESV02}, this is expected to encompass all admissible self-adjoint realizations of $H_{\alpha,S}$ in $L^2(\mathbb{R}^2)$. The focus is not on identifying minimal regularity hypotheses for $\mathbf{S}$, but rather on providing techniques which can be generalized to the case of multiple Aharonov-Bohm fluxes \cite{CFpre}. Building on the quadratic form construction, we further derive a natural convergence result showing that the Friedrichs realization of $H_{0,S} = H_{\alpha,S}\big|_{\alpha \,=\, 0}$ is the $\Gamma$-limit for $\alpha \to 0^{+}$ of the analogous realizations of $H_{\alpha,S}$ for $\alpha \in (0,1)$ (see Theorem \ref{thm:Gconv} and Corollary \ref{cor:ResCon}).
Let us finally mention that some of the results presented in this work might be of interest also for applications to anyonic particle models \cite{CLR17,CO18,LM77,O94,W82}, though this issue is not addressed directly here.
\vspace{0.2cm}

Throughout the paper we often refer to polar coordinates $(r,\theta) \!:\! \mathbf{R}^2 \setminus \{\mathbf{0}\} \to (0,+\infty) \times [0,2\pi)$ centered at $\mathbf{x} = \mathbf{0}$, and to the related angular averages of functions $f : \mathbb{R}^2 \to \mathbb{C}$, namely,
	\begin{equation*}
		\big\langle f \big\rangle(r) := {1 \over 2\pi r} \int_{\partial B_r(\mathbf{0})}\hspace{-0.3cm} d\Sigma_r\, f 
			= {1 \over 2\pi}\int_{0}^{2\pi}\!\!\!\! d\theta\, f\big(\mathbf{x}(r,\theta)\big)\,.
	\end{equation*}

\section{Main results}\vspace{-0.2cm}

\subsection{Self-adjoint realizations of $H_{\alpha,S}$}\vspace{-0.2cm}

Self-adjoint realizations in $L^2(\mathbb{R}^2)$ of the Schr\"odinger operator $H_{\alpha,S}$ are generally characterized as suitable extensions of the densely defined, symmetric operator $H_{\alpha,S}\!\! \upharpoonright\! C^{\infty}_c(\mathbb{R}^2 \!\setminus\! \{\mathbf{0}\})$. The simplest of such extensions is the Friedrichs one, to be denoted with $H_{\alpha,S}^{(F)}$ henceforth. This is obtained introducing the quadratic form\vspace{-0.1cm}
	\begin{equation*}
		Q_{\alpha,S}[\psi] \,= \int_{\mathbb{R}^2}\! d\mathbf{x}\;\big|(-i\nabla + \mathbf{A}_{\alpha} + \mathbf{S})\psi\big|^2\,, \qquad \mbox{for\, $\psi \in C^{\infty}_c(\mathbb{R}^2 \!\setminus\! \{\mathbf{0}\})$}\,,
	\end{equation*}
and considering its Friedrichs realization\vspace{-0.05cm}
	\begin{equation*}
		D\big[Q_{\alpha,S}^{(F)}\big] :=\, \overline{C^{\infty}_c(\mathbb{R}^2 \!\setminus\! \{\mathbf{0}\})}^{\,\|\,\cdot\,\|_{\alpha,S}}, \qquad Q^{(F)}_{\alpha,S}[\psi] = Q_{\alpha,S}[\psi]\,,
	\end{equation*}
where $\|\psi\|_{\alpha,S} \!:= \|\psi\|_2 + Q_{\alpha,S}[\psi]$. We recall the following Proposition from \cite{CF20}, to which we refer for the proof.

\begin{proposition}[Friedrichs realization]\label{prop:QHF}
Let $\alpha \in (0,1)$ and $\mathbf{S} \in L^{\infty}_{loc}(\mathbb{R}^2)$. Then:\\
i) The quadratic form $Q_{\alpha,S}^{(F)}$ is closed and non-negative on its domain, and\vspace{-0.1cm}
	\begin{equation}\label{eq:DQaSF}
		D\big[Q_{\alpha,S}^{(F)}\big] = \big\{\psi \!\in\! L^2(\mathbb{R}^2)\;\big|\;(-i \nabla + \mathbf{S})\psi \!\in\! L^2(\mathbb{R}^2)\,,\; \mathbf{A}_{\alpha}\psi \!\in\! L^2(\mathbb{R}^2) \big\}\,.\vspace{-0.1cm}
	\end{equation}
Moreover, any $\psi \!\in\! D\big[Q_{\alpha,S}^{(F)}\big]$ fulfills \vspace{-0.1cm}
	\begin{equation}\label{eq:asypsiQF}
		\lim_{r \to 0^{+}} \left\langle |\psi|^2 \right\rangle(r) = 0\,, \qquad\quad
		\lim_{r \to 0^{+}} r^2 \left\langle |\partial_r\psi|^2 \right\rangle(r) = 0\,. \vspace{-0.1cm}
	\end{equation}
ii) The unique self-adjoint operator $H_{\alpha,S}^{(F)}$ associated to $Q_{\alpha,S}^{(F)}$ is\vspace{-0.1cm}
	\begin{equation}\label{eq:DHaSF}
		D\big(H_{\alpha,S}^{(F)}\big) = \big\{\psi \!\in\! D\big[Q_{\alpha,S}^{(F)}\big] \;\big|\; H_{\alpha,S}\,\psi \!\in\! L^2(\mathbb{R}^2) \big\}\,, \qquad
		H_{\alpha,S}^{(F)}\, \psi = H_{\alpha,S}\, \psi .\vspace{0.15cm}
	\end{equation}
\end{proposition}

It is well known that the Friedrichs extension of a symmetric operator is the one with the smallest domain of self-adjointness. Accordingly, other self-adjoint realizations can only be obtained by suitable enlargements the domain. The standard approach to achieve this goal is to consider elements of the form $\psi = \phi_{\lambda} + q\, \mathcal{G}_{\lambda}$, where $\phi_{\lambda} \!\in\! D\big(H_{\alpha,S}^{(F)}\big)$, $q \!\in\! \mathbb{C}$ and $\mathcal{G}_{\lambda} \!\in\! L^2(\mathbb{R}^2)$, $\lambda \!>\! 0$, is a solution of the defect equation $(H_{\alpha,S} + \lambda^2)\mathcal{G}_{\lambda} = 0$ in $\mathbb{R}^2 \!\setminus\! \{\mathbf{0}\}$. A key obstacle in this construction is that, for generic choices of $\mathbf{S}$, an explicit expression of $\mathcal{G}_{\lambda}$ is not available. At the same time, we expect that the asymptotic behavior of $\mathcal{G}_{\lambda}$ near $\mathbf{x} = \mathbf{0}$ should not be affected by the perturbation $\mathbf{S}$, at least to leading order.

Trusting the latter surmise, we proceed to consider the solutions $G^{(k)}_{\lambda} \!\!\in\! L^2(\mathbb{R}^2)$, $k \!\in\! \{0,-1\}$, of the unperturbed defect equation \vspace{-0.1cm}
	\begin{equation}\label{eq:defGEq}
		\big((- i \nabla + \mathbf{A}_{\alpha})^2 + \lambda^2\big)G^{(k)}_{\lambda} \!= 0\,, \qquad \mbox{in\, $\mathbb{R}^2 \!\setminus\! \{\mathbf{0}\}$}\,.
	\end{equation}
%\pagebreak
The parameter $k$ coincides with the angular momentum number of the wavefunction $G^{(k)}_{\lambda}$. For this reason, we shall refer to $G^{(0)}_{\lambda}$ and $G^{(-1)}_{\lambda}$, respectively, as the \emph{s-wave} and \emph{p-wave Green functions}.\footnote{By decomposition in angular harmonics, it appears that \eqref{eq:defGEq} admits non-trivial solutions only for $k = 0$ and $k = -1$.}\\
Using angular coordinates and writing $K_{\nu}$ for the modified Bessel function of second kind (\emph{a.k.a.} Macdonald function), their explicit expressions are given by
	\begin{equation}\label{eq:G01exp}
		G^{(k)}_{\lambda}(r,\theta) \equiv G^{(k)}_{\lambda}\big(\mathbf{x}(r,\theta)\big) 
		= \lambda^{|k+\alpha|}\,K_{|k+\alpha|}(\lambda r)\,e^{i k \theta}, \quad\;
		\mbox{for\, $k \!\in\!\{0,-1\}$}\,. 
	\end{equation}
For later reference, let us mention that (see \cite[Eq. 6.521.3]{GR07})
	\begin{equation}\label{eq:GjL2}
		\big\|G^{(k)}_{\lambda}\big\|_{2}^{2} = {\pi^2\, |k+\alpha| \over \sin(\pi \alpha)}\,\lambda^{2|k+\alpha| - 2}\,, \qquad\; \mbox{for\, $k\!\in\!\{0,-1\}$}\,.
	\end{equation}
To say more, for $r \to 0^{+}$ there holds (see \cite[\S 10.31]{OLBC10})
\begin{align}
	G^{(k)}_{\lambda}\!(r,\theta) = \left[{\Gamma\big( |k+\alpha| \big) \over 2^{1-|k+\alpha|}}\,{1 \over r^{|k+\alpha|}} + {\Gamma\big(\!-|k+\alpha|\big) \over 2^{1 + |k+\alpha|}}\, \lambda^{2|k+\alpha|}\,r^{|k+\alpha|} + \mathcal{O}\big(r^{2-|k+\alpha|}\big)\right]e^{i k \theta} , \label{eq:G01asy}
\end{align}

So far, we made no assumption concerning the regularity of $\mathbf{S}$ near the origin, where the Aharonov-Bohm potential $\mathbf{A}_{\alpha}$ is singular. We henceforth require that
	\begin{equation}\label{eq:SLip}
		\mbox{$\mathbf{S}\in L^{\infty}_{\mbox{{\scriptsize loc}}}(\mathbb{R}^2,\mathbb{R}^2)$\, is Lipschitz continuous at\, $\mathbf{x} = \mathbf{0}$}\,.
	\end{equation}
Without loss of generality we also fix the Coulomb gauge, which entails
	\begin{equation}\label{eq:Coul}
		\nabla \cdot \mathbf{S} = 0\,.
	\end{equation}
Under the above hypotheses, for $\lambda > 0$ we consider trial functions of the form
	\begin{equation}\label{eq:psipert}
		\psi = \phi_{\lambda} + e^{-i\, \mathbf{S}(\mathbf{0}) \cdot \mathbf{x}}\,\chi \sum_{k \in \{0,-1\}}  q^{(k)}\, G^{(k)}_{\lambda}\,, 
	\end{equation}
where $\phi_{\lambda} \!\in\! D\big[Q_{\alpha,S}^{(F)}\big]$, $q^{(k)} \!\in\! \mathbb{C}$ for $k \!\in\! \{0,-1\}$ and $\chi \!: \mathbb{R}^2 \!\to\! [0,1]$ is a smooth cut-off function fulfilling
	\begin{equation}\label{eq:Hypchi}
		\chi \!\in\! C^{2}_c(\mathbb{R}^2)\,, \qquad\quad
		\chi(\mathbf{x}) \!\equiv\! 1 \quad \mbox{for any\, $\mathbf{x} \!\in\! B_{a}(\mathbf{0})$, for some\, $a \!>\! 0$}\,.
	\end{equation}
The latter cut-off is necessary to include in the present analysis the case of perturbations $\mathbf{S}$ which are unbounded at infinity, comprising especially configurations with magnetic traps. For the sake of simplicity, we assume $\chi$ to be radial, \emph{i.e.},
	\begin{equation}\label{eq:chirad}
		\chi(\mathbf{x}) \equiv \chi\big(|\mathbf{x}|\big) \equiv \chi(r)\,.
	\end{equation}

\begin{remark} 
One could fix $\mathbf{S}(\mathbf{0}) \!=\! \mathbf{0}$, on top of the Coulomb gauge \eqref{eq:Coul}. This would make the phase factor in \eqref{eq:psipert} irrelevant and would even allow to abridge some of the expressions to be derived in the following. Yet, in this work we choose not to fix the value of $\mathbf{S}$ at $\mathbf{x} \!=\! \mathbf{0}$ in order to exhibit a construction which can be generalized to the case of multiple fluxes with a moderate effort \cite{CFpre}.
With the same objective in mind, we stick to the Lipschitz condition in \eqref{eq:SLip}, though most of the results in this work are still valid requiring just some H\"older regularity of $\mathbf{S}$ at the origin.
\end{remark}

A heuristic evaluation of the expectation value $\langle \psi | H_{\alpha,S}\, \psi \rangle$ for functions $\psi$ like \eqref{eq:psipert} suggests the educated guess
	\begin{align}
		& Q_{\alpha,S}^{(\beta)}[\psi] \nonumber
			:= Q_{\alpha,S}^{(F)}[\phi_{\lambda}] - \lambda^2\,\|\psi\|_2^{2} + \lambda^2\,\|\phi_{\lambda}\|_{2}^{2} \\
		& + 2\! \sum_{k \in \{0,-1\}}\!\! \mbox{Re} \left[ q^{(k)} \left( 2 \left\langle (- i \nabla + \mathbf{A}_{\alpha}) \phi_{\lambda} \left|\, e^{- i\, \mathbf{S}(\mathbf{0}) \cdot \mathbf{x}} \big(\big(\mathbf{S} - \mathbf{S}(\mathbf{0})\big)\chi - i \nabla\chi \big) G^{(k)}_{\lambda} \!\right.\right\rangle \right.\right.\nonumber \\
		& \hspace{0.5cm} \left.\left. + \, \left\langle \phi_{\lambda} \left|\, e^{-i\, \mathbf{S}(\mathbf{0}) \cdot \mathbf{x}} \big[\big(\mathbf{S} - \mathbf{S}(\mathbf{0})\big)^2 \chi + 2\mathbf{S}(\mathbf{0}) \!\cdot\! \big(\big(\mathbf{S} - \mathbf{S}(\mathbf{0})\big)\chi - i \nabla\chi \big) + \Delta \chi \big] G^{(k)}_{\lambda}\! \right.\right\rangle \right) \right] \nonumber\\
		& + \sum_{k,k' \in \{0,-1\}}\!\! \overline{q^{(k)}}\, q^{(k')}\! \left[ \beta_{kk'} + {\pi^2 \over \sin(\pi \alpha)}\, \lambda^{2 |k+\alpha|}\, \delta_{kk'} + \Xi_{kk'}(\lambda) \right] , \label{eq:QaSb}
\end{align}
where we have introduced the $2\!\times\!2$ complex Hermitian matrix $\beta = (\beta_{kk'})$, labeling the quadratic form, and we have set
	\begin{align}
		& \Xi_{kk'}(\lambda) := \left\langle \chi\, G^{(k)}_{\lambda} \left|\, \big[ \big(\mathbf{S} - \mathbf{S}(\mathbf{0})\big)^2\! + 2 \big(\mathbf{S} - \mathbf{S}(\mathbf{0})\big) \!\cdot\! \mathbf{A}_{\alpha}\big] \chi\,G^{(k')}_{\lambda} \!\right.\right\rangle  \nonumber \\
		& \hspace{1.5cm} +\, \big\|(\nabla \chi)\, G^{(k)}_{\lambda}\big\|_{2}^{2}\;\delta_{kk'} +  2 \left\langle \chi \, G^{(k)}_{\lambda} \left|\, \big(\mathbf{S} - \mathbf{S}(\mathbf{0})\big) \!\cdot\! (- i \nabla)\big(\chi\, G^{(k')}_{\lambda}\big)\! \right.\right\rangle . \label{eq:Xidef}
	\end{align}
The expression \eqref{eq:QaSb} was derived integrating by parts and deliberately discarding some contributions, an operation to be justified a posteriori in the proof of Theorem \ref{thm:QasB}. We also used that $\mathbf{A}_{\alpha} \!\cdot\! \nabla\chi = 0$, since $\chi$ is radial (see \eqref{eq:chirad}), and the identity
	\begin{equation}\label{eq:GkGkrad}
		\big\langle G^{(k)}_{\lambda} \,\big| \,\eta\,G^{(k')}_{\lambda} \big\rangle = \big\langle G^{(k)}_{\lambda} \,\big| \,\eta\,G^{(k)}_{\lambda} \big\rangle\;\delta_{k k'}\,, \quad\; \mbox{for any radial\, $\eta \!:\! \mathbb{R}^2\! \to\! \mathbb{R}$}\,.
	\end{equation}

\begin{remark}\label{rem:XiHerm}
For any $\lambda \!>\! 0$, the matrix $\Xi_{kk'}(\lambda)$, $k,k' \!\in\! \{0,-1\}$, defined by \eqref{eq:Xidef} is itself Hermitian. This feature is evident for the first two addenda in \eqref{eq:Xidef}, given that $\mathbf{S},\mathbf{A}_{\alpha}$ and $\chi$ are real-valued. As regards the last addendum in \eqref{eq:Xidef}, integrating by parts and checking that the boundary contribution vanishes (recall \eqref{eq:G01asy} and \eqref{eq:SLip}), we have in fact
	\begin{align*}
		\left\langle \chi \, G^{(k)}_{\lambda} \left|\, \big(\mathbf{S} - \mathbf{S}(\mathbf{0})\big) \!\cdot\! (- i \nabla)\big(\chi\, G^{(k')}_{\lambda}\big)\! \right.\right\rangle
		& = \left\langle (- i \nabla)\big(\chi \, G^{(k)}_{\lambda}\big) \left|\, \big(\mathbf{S} - \mathbf{S}(\mathbf{0})\big) \chi\, G^{(k')}_{\lambda}\! \right.\right\rangle \\
		& = \overline{\left\langle \chi \, G^{(k')}_{\lambda} \left|\, \big(\mathbf{S} - \mathbf{S}(\mathbf{0})\big) \!\cdot\! (- i \nabla)\big(\chi\, G^{(k)}_{\lambda}\big)\! \right.\right\rangle}\,.
	\end{align*}
Besides, building on the fact that $G^{(0)}_{\lambda}$ is real-valued, c.f.\! \eqref{eq:G01exp}, it can be inferred that
	\begin{equation}
		\left\langle \chi \, G^{(k)}_{\lambda} \left|\, \big(\mathbf{S} - \mathbf{S}(\mathbf{0})\big) \!\cdot\! (- i \nabla)\big(\chi\, G^{(k')}_{\lambda}\big)\! \right.\right\rangle = 0\,, \qquad \mbox{for\, $k \!=\! k' \!=\! 0$}\,.
	\end{equation}
\end{remark}

A family of admissible singular perturbations of the Friedrichs realization is described by the upcoming Theorem \ref{thm:QasB} and Corollary \ref{cor:HasB}.

\begin{theorem}[Quadratic forms for singular perturbations]\label{thm:QasB}
Let $\alpha \in (0,1)$ and $\mathbf{S} \in L^{\infty}_{loc}(\mathbf{R}^2)$ be Lipschitz continuous at $\mathbf{x} = \mathbf{0}$, with $\nabla \cdot \mathbf{S} = 0$. Then, for any Hermitian matrix $\beta = (\beta_{kk'})$, $k,k'\!\in\!\{0,-1\}$, the quadratic form $Q_{\alpha,S}^{(\beta)}$ defined in \eqref{eq:QaSb} satisfies the following:\\
i) It is well-posed on the domain 
	\begin{align}
		& D\big[Q_{\alpha,S}^{(\beta)}\big] := \Big\{\, \psi = \phi_{\lambda} + e^{-i\, \mathbf{S}(\mathbf{0}) \cdot \mathbf{x}}\,\chi\, \mbox{$\sum_{k \in \{0,-1\}}$}  q^{(k)}\, G^{(k)}_{\lambda} \in L^2(\mathbb{R}^2) \quad \mbox{s.t.} \nonumber \\
		& \hspace{1.3cm} \phi_{\lambda}\!\in\! D\big[Q_{\alpha,S}^{(F)}\big],\,\lambda \!>\! 0\,,\;\;\mbox{$\chi$ fulfills \eqref{eq:Hypchi}\! \eqref{eq:chirad}}\,, \;\; q^{(k)} \!\in\! \mathbb{C}\,, k \!\in\! \{0,-1\} \Big\}\,.
			\label{eq:domQb}
	\end{align}
ii) It is independent of $\lambda \!>\! 0$ and of the cut-off $\chi$, provided that \eqref{eq:Hypchi}\! \eqref{eq:chirad} hold true.\\
iii) It is closed and bounded from below on the domain \eqref{eq:domQb}.
\end{theorem}

\begin{corollary}[Self-adjoint realizations for singular perturbations]\label{cor:HasB}
Assume the hypotheses of Theorem \ref{thm:QasB} to hold. Then, for any $2 \times 2$ Hermitian matrix $\beta$, the self-adjoint operator $H_{\alpha,S}^{(\beta)}$ associated to the quadratic form $Q_{\alpha,S}^{(\beta)}$ is given by
	\begin{align}
		& D\big(H_{\alpha,S}^{(\beta)}\big) = \Big\{\, \psi = \phi_{\lambda} + e^{-i\, \mathbf{S}(\mathbf{0}) \cdot \mathbf{x}} \chi \mbox{$\sum_{k \in \{0,-1\}}$}  q^{(k)} G^{(k)}_{\lambda} \in D\big[Q_{\alpha,S}^{(\beta)}\big] \;\quad \mbox{s.t.} \nonumber \\
		& \hspace{1.9cm} \phi_{\lambda} \in D\big(H_{\alpha,S}^{(F)}\big) \quad\; \mbox{and} \label{eq:DHaSbe} \\
		& \hspace{1.9cm} {2^{1-|k+\alpha|} \over \Gamma\big( |k+\alpha| \big)} \sum_{k' \in \{0,-1\}}\! q^{(k')} \bigg(\,\beta_{kk'} + {\pi^2 \lambda^{2 |k+\alpha|} \over \sin(\pi \alpha)}\,\delta_{kk'} \bigg) \nonumber \\
		& \hspace{2.3cm} = \lim_{r \to 0^{+}}\! {|k\!+\!\alpha|\, \langle e^{- i k \theta}\! \phi_{\lambda} \rangle(r) + r \langle e^{- i k \theta}\! \partial_{r}\phi_{\lambda} \rangle(r) \over r^{|k+\alpha|}}\,, \;\; \mbox{for $k\!\in\!\{0,-1\}$}\bigg\}\,; \nonumber 
	\end{align}
	\begin{align}
		& \big(H_{\alpha,S}^{(\beta)} + \lambda^2\big) \psi = \big(H_{\alpha,S}^{(F)} + \lambda^2\big) \phi_{\lambda} \nonumber  \\
		& \hspace{2.4cm} +\!\! \sum_{k \in \{0,-1\}} \!\!\!q^{(k)} e^{- i\, \mathbf{S}(\mathbf{0}) \cdot \mathbf{x}} \!\left[  2 \Big(\big(\mathbf{S} - \mathbf{S}(\mathbf{0})\big)\chi - i \nabla\chi \Big) \!\cdot\! \big(\!- i \nabla + \mathbf{A}_{\alpha}\big) G^{(k)}_{\lambda} \right.\nonumber \\
		& \hspace{3.cm} \left. + \, \Big(\big(\mathbf{S} - \mathbf{S}(\mathbf{0})\big)^2 \chi + 2\big(\mathbf{S} - \mathbf{S}(\mathbf{0})\big)\!\cdot\!(- i \nabla\chi) - \Delta \chi \Big) G^{(k)}_{\lambda}\right] . \label{eq:HaSbe}
	\end{align}
The set of operators $H_{\alpha,S}^{(\beta)}$, $\beta$ Hermitian, identifies a family of self-adjoint extensions of $H_{\alpha,S}\! \upharpoonright\! C^{\infty}_c(\mathbb{R}^2 \!\setminus\! \{\mathbf{0}\})$ in $L^2(\mathbb{R}^2)$, labeled by four real parameters.\footnote{Notice that $2\!\times\! 2$ complex Hermitian matrices form a 4-dimensional real vector space.} If $\mathbf{S} \!\in\! L^{\infty}(\mathbb{R}^2)$, this family comprises all admissible self-adjoint realizations of $H_{\alpha,S}$ in $L^2(\mathbb{R}^2)$.
\end{corollary}

\begin{remark}
The characterization \eqref{eq:DHaSbe} of the operator domain is quite standard. Especially, it incorporates boundary conditions relating the ``charges'' $q^{(k)}$ to the asymptotic behavior of the ``regular part'' $\phi_{\lambda}$ close to $\mathbf{x} \!=\! \mathbf{0}$. Considering that the matrix $\beta_{kk'} + {\pi^2 \lambda^{2 |k+\alpha|} \over \sin(\pi \alpha)}\,\delta_{kk'}$ is certainly invertible for $\lambda$ large enough, it is always possible to derive an explicit expression for $q^{(k)}$ in terms of boundary values of $\phi_{\lambda}$.
Let us also stress that, in agreement with our expectations, only the leading order term of the asymptotic expansion \eqref{eq:G01asy} for $G_{\lambda}^{(k)}$ is relevant here (\emph{cf.} the proof of Corollary \ref{cor:HasB}).
\end{remark}

\begin{remark}
The Hermitian matrix $\beta$ labeling the self-adjoint operator $H_{\alpha,S}^{(\beta)}$ only appears in the boundary conditions for $D\big(H_{\alpha,S}^{(\beta)}\big)$. In this sense, it parametrizes a singular interaction affecting just the \emph{s-wave} and \emph{p-wave} modes of the wave-functions. In \cite{CF20} attention was restricted to pure \emph{s-wave} perturbations, corresponding to $\beta_{kk'} \!=\! b\, \delta_{k,0}\, \delta_{k',0}$ with $b \!\in\! \mathbb{R}$. Here we also include pure \emph{p-wave} perturbations, as well as mixed interactions coupling \emph{s-wave} and \emph{p-wave} modes.
\end{remark}

\begin{remark}
The Friedrichs realization is recovered for $q^{(0)} \!=\! q^{(-1)} \!=\! 0$. This condition formally corresponds to fixing $\beta_{kk'} \!=\! (+ \infty)\, \delta_{kk'}$. The characterization \eqref{eq:DHaSbe} of $D\big(H_{\alpha,S}^{(F)}\big)$ states explicitly the boundary condition\vspace{-0.1cm}
	\begin{equation*}
		\lim_{r \to 0^{+}}\! {|k\!+\!\alpha|\, \langle e^{- i k \theta}\! \phi_{\lambda} \rangle(r) + r \langle e^{- i k \theta}\! \partial_{r}\phi_{\lambda} \rangle(r) \over r^{|k+\alpha|}} = 0\,, \qquad \mbox{for $k\!\in\!\{0,-1\}$}\,,
	\end{equation*}
which is otherwise concealed in the requirement $H_{\alpha,S}^{(F)}\, \phi_{\lambda} \!\in\! L^2(\mathbb{R}^2)$ of \eqref{eq:DQaSF}.
\end{remark}

\begin{remark}
The action of the operator described in \eqref{eq:HaSbe} is somehow unorthodox. Making reference to the standard theory of self-adjoint extensions, one would rather expect the simpler relation $\big(H_{\alpha,S}^{(\beta)} + \lambda^2\big) \psi \!=\! \big(H_{\alpha,S}^{(F)} + \lambda^2\big) \phi_{\lambda}$. The expressions in the last two lines of \eqref{eq:HaSbe} are in fact necessary corrections, produced by the use of surrogates in place of true defect functions for $H_{\alpha,S}$.
\end{remark}

\begin{remark}
Electrostatic potentials regular enough at the Aharonov-Bohm singularity could be easily incorporated in the construction provided here. We omit the discussion of this further development for the sake of brevity.
\end{remark}

\subsection{$\Gamma$-convergence for the Friedrichs Hamiltonian}\vspace{-0.1cm}

Consider now a regime where the Aharonov-Bohm flux is negligible, in suitable units, compared to the external magnetic perturbation or to the angular momentum of the particle. In this context, the dynamics of the particle should be properly described by some self-adjoint realization in $L^2(\mathbb{R}^2)$ of the Schr\"odinger operator
	\begin{equation}
		H_{0,S} \equiv H_{\alpha,S}\big|_{\alpha \, =\, 0} \,=\, \big(-i \nabla + \mathbf{S}\big)^2.
	\end{equation}
At the same time, due to the local singularity at $\mathbf{x} \!=\! \mathbf{0}$ of the Aharonov-Bohm potential $\mathbf{A}_{\alpha}$, establishing the convergence $H_{\alpha,S} \!\to\! H_{0,S}$ for $\alpha \!\to\! 0^{+}$ (in any reasonable topology) is not a plain task. Building on the quadratic form approach described in the previous subsection, we present hereafter a result based on the classical notion of $\Gamma$\! - convergence \cite{Br02,DM93}.

For the sake of simplicity, let us assume that\footnote{Notice the similarity with \eqref{eq:SLip}. Here we are making a stronger requirement: $\mathbf{S}$ must be uniformly bounded on the whole space $\mathbb{R}^2$, not just on compact subsets of it. This excludes magnetic traps.}
	\begin{equation}\label{eq:SLipinf}
		\mbox{$\mathbf{S}\in L^{\infty}(\mathbb{R}^2,\mathbb{R}^2)$\, is Lipschitz continuous at\, $\mathbf{x} = \mathbf{0}$}\,.
	\end{equation}
Besides, we restrict the attention to the Friedrichs Hamiltonian $H_{\alpha,S}^{(F)}$ of Proposition \ref{prop:QHF}, postponing the discussion of the singular perturbations $H_{\alpha,S}^{(\beta)}$ characterized in Theorem \ref{thm:QasB} and Corollary \ref{cor:HasB} to future investigations.

Let us consider the Friedrichs quadratic form $Q_{\alpha,S}^{(F)}$ (see Proposition \ref{prop:QHF}) and extend it to the whole Hilbert space $L^2(\mathbb{R}^2)$ setting
	\begin{equation}
		Q_{\alpha,S}^{(F)}[\psi] := \left\{\,\begin{array}{ll}
			\displaystyle{ \big\|(-i \nabla + \mathbf{A}_{\alpha} + \mathbf{S})\psi\big\|_{2}^{2}}
				& \quad\displaystyle{\mbox{if\, $\psi\!\in\! D\big[Q_{\alpha,S}^{(F)}\big]$}\,;} \vspace{0.15cm} \\
			\displaystyle{+ \infty}
				& \quad\displaystyle{\mbox{if\, $\psi\!\in\! L^2(\mathbb{R}^2) \!\setminus\! D\big[Q_{\alpha,S}^{(F)}\big]$}\,.}
		\end{array} \right. \label{eq:QaSext}
	\end{equation}
Notice that, under the hypothesis \eqref{eq:SLipinf}, the identity \eqref{eq:DQaSF} in Proposition \ref{prop:QHF} reduces to 
	\begin{equation*}
		D\big[Q_{\alpha,S}^{(F)}\big] = \big\{\psi \!\in\! H^1(\mathbb{R}^2)\;\big|\; \mathbf{A}_{\alpha} \psi \!\in\! L^2(\mathbb{R}^2)\big\}\,.
	\end{equation*}
In a similar fashion, for $\alpha = 0$ we put
	\begin{equation}
		Q_{0,S}^{(F)}[\psi] := \left\{\,\begin{array}{ll}
			\displaystyle{\big\|(-i \nabla + \mathbf{S})\psi\big\|_{2}^{2}}
				& \quad\displaystyle{\mbox{if\, $\psi\!\in\! D\big[Q_{0,S}^{(F)}\big] \equiv H^1(\mathbb{R}^2)$}\,;} \vspace{0.2cm} \\
			\displaystyle{+ \infty}
				& \quad\displaystyle{\mbox{if\, $\psi\!\in\! L^2(\mathbb{R}^2) \!\setminus\! H^1(\mathbb{R}^2)$}\,.}
		\end{array} \right. \label{eq:Q0Sext}
	\end{equation}
For later reference, let us mention that the self-adjoint operators associated to the above quadratic forms are respectively given by (\emph{cf.} \eqref{eq:DHaSF})
	\begin{gather*}
		D\big(H_{\alpha,S}^{(F)}\big) = \big\{\psi \!\in\! H^1(\mathbb{R}^2) \;\big|\; \mathbf{A}_{\alpha} \psi, H_{\alpha,S}\,\psi \!\in\! L^2(\mathbb{R}^2) \big\}\,, \qquad
			H_{\alpha,S}^{(F)}\, \psi = H_{\alpha,S}\, \psi\,;\\
		D\big(H_{0,S}^{(F)}\big) = H^2(\mathbb{R}^2)\,, \qquad
			H_{0,S}^{(F)}\, \psi = H_{0,S}\, \psi\,.
\end{gather*}

\begin{theorem}\label{thm:Gconv}
Let $\mathbf{S} \in L^{\infty}(\mathbf{R}^2)$ be Lipschitz continuous at $\mathbf{x} = \mathbf{0}$, with $\nabla \cdot \mathbf{S} = 0$, and $\{\alpha_n\}_{n\,\in\,\mathbb{N}} \!\subset\! (0,1)$ be any sequence such that $\alpha_n \to 0$\, for $n \to +\infty$. Then, the family of quadratic forms $Q_{\alpha_n,S}^{(F)}$ $\Gamma$-converges to $Q_{0,S}^{(F)}$, that is:\\
i) \emph{Lower bound inequality.} For every sequence $\{\psi_{\alpha_n}\}_{n\,\in\,\mathbb{N}} \!\subset\! L^2(\mathbb{R}^2)$ such that $\psi_{\alpha_n} \!\to \psi_{0} \!\in\! L^2(\mathbb{R}^2)$ as $n \to +\infty$, there holds
	\begin{equation}\label{eq:lowerb}
		Q_{0,S}^{(F)}[\psi_0] \leqslant \liminf_{n \to +\infty}\, Q_{\alpha_n,S}^{(F)}[\psi_{\alpha_n}]\,.
	\end{equation}
ii) \emph{Upper bound inequality.} For every $\psi_0\!\in\!L^2(\mathbb{R}^2)$ there exists a sequence $\{\psi_{\alpha_n}\}_{n\,\in\,\mathbb{N}} \!\subset\! L^2(\mathbb{R}^2)$ such that $\psi_{\alpha_n} \!\to \psi_{0}$ as $n \to +\infty$ and
	\begin{equation}\label{eq:upperb}
		Q_{0,S}^{(F)}[\psi_0] \geqslant \limsup_{n \to +\infty}\, Q_{\alpha_n,S}^{(F)}[\psi_{\alpha_n}]\,.
	\end{equation}
\end{theorem}

From the previous theorem and classical results on $\Gamma$\! - convergence \cite[\S 13]{DM93}, we readily deduce the following.

\begin{corollary}\label{cor:ResCon}
Under the same assumptions of Theorem \ref{thm:Gconv}, the family of operators $H_{\alpha_n,S}^{(F)}$ converges to $H_{0,S}^{(F)}$ in strong resolvent sense for $n \to +\infty$. More precisely, for any $z \!\in\! \mathbb{C} \!\setminus\! [0,+\infty)$ and any $\psi\!\in\!L^2(\mathbb{R}^2)$, there holds
	\begin{equation}
		\left\| \big( H_{\alpha_n,S}^{(F)} - z \big)^{-1} \psi - \big( H_{0,S}^{(F)} - z \big)^{-1} \psi \right\|_2 \,\xrightarrow{n \to +\infty}\; 0\,.
	\end{equation}
\end{corollary}

\begin{remark}
The requirement $z \!\in\! \mathbb{C} \!\setminus\! [0,+\infty)$ in Corollary \ref{cor:ResCon} matches the elementary inclusions $\sigma\big(H_{\alpha_n,S}^{(F)}\big) \!\subset\! [0,+\infty)$ and $\sigma\big(H_{0,S}^{(F)}\big) \!\subset\! [0,+\infty)$.
\end{remark}

\begin{remark}
In the pure Aharonov-Bohm configuration, with $\mathbf{S} = \mathbf{0}$, it should be possible to infer strong resolvent convergence for $\alpha \to 0$ even by direct computations, starting from the explicit expression for the integral kernel of the resolvent operator derived in \cite{AT98}. This alternative approach would however involve a rather complicate analysis, relying on non-elementary regularity features of the Bessel functions with respect to their order and further demanding non-trivial exchanges of limits and integrations. On top of that, the $\Gamma$\! - convergence method considered in this work appears to be more flexible. Especially, it should be possible to adapt it to multiple fluxes configurations with not too much effort \cite{CFpre}.
\end{remark}

\begin{remark}
Despite being quite natural, the results derived in Theorem \ref{thm:Gconv} and Corollary \ref{cor:ResCon} are not completely obvious, especially if one considers the topology of the underlying space domains. In fact, the Aharonov-Bohm configuration ($\alpha \neq 0$) refers to the domain $\mathbb{R}^2 \!\setminus\! \{\mathbf{0}\}$, with first homotopy group given by $\mathbb{Z}$, while the setting with no singular flux ($\alpha = 0$) corresponds to the plane $\mathbb{R}^2$, with trivial homotopy group.
\end{remark}

\section{Proofs}

Let us recall that Theorem \ref{thm:QasB} and Corollary \ref{cor:HasB} rely on the hypothesis \eqref{eq:SLip} for $\mathbf{S}$, demanding $\mathbf{S}$ to be locally uniformly bounded and Lipschitz continuous at $\mathbf{x} = \mathbf{0}$.

\begin{proof}[Theorem \ref{thm:QasB}] Each of the statements \emph{i) - iii)} can be derived adapting some related arguments from \cite{CF20}. Throughout the proof, $\mathbf{1}_{\chi}$ is the indicator function of the support of $\chi$ and $c \!\equiv\! c(\alpha,\mathbf{S})$ is a suitable positive constant independent of $\lambda$, which may vary from line to line.

{\sl i)} Upon identifying $(-i\nabla + \mathbf{A}_{\alpha})\phi_{\lambda}$ with $\mathbf{1}_{\chi} (-i\nabla + \mathbf{A}_{\alpha})\phi_{\lambda}$ in \eqref{eq:QaSb}, all parings in \eqref{eq:QaSb} \eqref{eq:Xidef} are well-defined inner products in $L^2(\mathbb{R}^2)$. To account for this claim, firstly note that $(-i\nabla + \mathbf{A}_{\alpha})\phi_{\lambda} \!\in \!L^2_{\mbox{{\scriptsize loc}}}(\mathbb{R}^2)$ for any $\phi_{\lambda} \!\in\! D\big[Q_{\alpha,S}^{(F)}\big]$, see \eqref{eq:DQaSF}. Secondly, recall that $G^{(k)}_{\lambda} \!\in\!L^2(\mathbb{R}^2)$ for $k\!\in\!\{0,-1\}$, see \eqref{eq:GjL2}. Hypotheses \eqref{eq:SLip} and \eqref{eq:Hypchi} further grant the uniform boundedness of $\nabla\chi$, $\Delta \chi$, $\big(\mathbf{S} - \mathbf{S}(\mathbf{0})\big)\chi$ and $\big(\mathbf{S} - \mathbf{S}(\mathbf{0})\big) \cdot \mathbf{A}_{\alpha} \chi$. In view of the basic relation $\big|\nabla \big(\chi G^{(k)}_{\lambda}\big)\big| \!\leqslant\! {c \over |\mathbf{x}|}\, \chi G^{(k)}_{\lambda}$, $k\!\in\!\{0,-1\}$, the same hypotheses also ensure that $\big(\mathbf{S} - \mathbf{S}(\mathbf{0})\big) \!\cdot\! (- i \nabla)\big(\chi\, G^{(k)}_{\lambda}\big) \!\in\! L^2(\mathbb{R}^2)$.

{\sl ii)} Let us show that the form is independent of $\lambda \!>\! 0$. To this purpose, fix $\lambda_1 \!\neq\! \lambda_2$ and consider, for any $\psi \!\in\! D\big[Q^{(\beta)}_{\alpha,S}\big]$, the two alternative representations $\psi \!=\! \phi_{\lambda_1}\! + e^{-i \mathbf{S}(\mathbf{0}) \cdot \mathbf{x}} \chi \sum_{k \in \{0,-1\}} q^{(k)} G_{\lambda_1}^{(k)}$ and $\psi \!=\! \phi_{\lambda_2}\! + e^{-i \mathbf{S}(\mathbf{0}) \cdot \mathbf{x}} \chi \sum_{k \in \{0,-1\}} q^{(k)} G_{\lambda_2}^{(k)}\!$. It is easy to check that $\chi \big( G_{\lambda_2}^{(k)}\! - G_{\lambda_1}^{(k)}\big) \!\in\! D\big[Q_{\alpha,S}^{(F)}\big]$ for $k \!\in\! \{0, -1\}$ (see \eqref{eq:DQaSF} and \eqref{eq:G01asy}).\linebreak This ensures that the ``charges'' $q^{(k)}$ are independent of $\lambda$, and further entails $\phi_{\lambda_1} \!\!=\! \phi_{\lambda_2}\! + e^{-i \mathbf{S}(\mathbf{0}) \cdot \mathbf{x}} \chi \sum_{k \in \{0,-1\}} q^{(k)} (G_{\lambda_2}^{(k)}\! - G_{\lambda_1}^{(k)})$.
Taking these facts into account and exploiting the identity \eqref{eq:GkGkrad}, with a number of integrations by parts we obtain
{\small 
\begin{align}
	& Q_{\alpha,S}^{(\beta)}\left[\phi_{\lambda_1}\! + e^{-i \mathbf{S}(\mathbf{0}) \cdot \mathbf{x}} \chi \mbox{$\sum_{k \in \{0,-1\}}$} q^{(k)} G_{\lambda_1}^{(k)}\right] \nonumber \\
	& = \langle \phi_{\lambda_2} | (-i \nabla + \mathbf{A}_{\alpha} + \mathbf{S})^2 \phi_{\lambda_2} \rangle 
		- \lambda_2^2\,\|\psi\|_2^{2} + \lambda_2^2 \,\|\phi_{\lambda}\|_{2}^{2} \nonumber \\
	& \quad + 2\!\sum_{k \in \{0,-1\}}\!\! \mbox{Re} \left[ q^{(k)} \left( 2 \left\langle (- i \nabla + \mathbf{A}_{\alpha}) \phi_{\lambda_2} \left|\, e^{-i \mathbf{S}(\mathbf{0}) \cdot \mathbf{x}} \big[\big(\mathbf{S} - \mathbf{S}(\mathbf{0})\big)\chi - i \nabla\chi \big] G^{(k)}_{\lambda_2} \right.\right\rangle \right. \right. \nonumber \\
		& \hspace{0.8cm} \left. \left. + \left\langle \phi_{\lambda_2} \left|\, e^{-i \mathbf{S}(\mathbf{0}) \cdot \mathbf{x}}  \big[ \big(\mathbf{S} - \mathbf{S}(\mathbf{0})\big)^2 \chi + 2\, \mathbf{S}(\mathbf{0}) \!\cdot\! \big(\big(\mathbf{S} - \mathbf{S}(\mathbf{0})\big)\chi - i \nabla\chi \big) + \Delta \chi \big] G^{(k)}_{\lambda_2} \right.\right\rangle \right) \right] \nonumber \\
	& \quad +\! \sum_{k,k' \in \{0,-1\}}\!\! \overline{q^{(k)}}\, q^{(k')} \bigg[ \beta_{kk'} + {\pi^2 \over \sin(\pi \alpha)}\,\lambda_2^{2 |k+\alpha|}\, \delta_{ij} +\; 2 \left\langle \chi \, G^{(k)}_{\lambda_2} \left|\, \big(\mathbf{S} - \mathbf{S}(\mathbf{0})\big) \!\cdot\! (- i \nabla)\big(\chi\, G^{(k')}_{\lambda_2}\big)\! \right.\right\rangle \nonumber \\
		& \hspace{2.2cm} + \left\langle \chi G^{(k)}_{\lambda_2} \left|\, \big[ \big(\mathbf{S} - \mathbf{S}(\mathbf{0})\big)^2\! + 2 \big(\mathbf{S} - \mathbf{S}(\mathbf{0})\big) \!\cdot\! \mathbf{A}_{\alpha}\big] \chi G^{(k')}_{\lambda_2} \!\right.\right\rangle + \big\|(\nabla \chi)\, G^{(k)}_{\lambda_2}\big\|_{2}^{2}\;\delta_{kk'} \bigg] \nonumber \\
%%%%%%%%%%%
	& \quad +\! \sum_{k \in \{0,-1\}}\! \big|q^{(k)}\big|^2\, \bigg[{\pi^2 \over \sin(\pi \alpha)} \left(\lambda_1^{2 |k+\alpha|} \!-\! \lambda_2^{2 |k+\alpha|}\right) + (\lambda_2^2 - \lambda_1^2) \big\langle \chi G_{\lambda_1}^{(k)} \big|\, \chi  G_{\lambda_2}^{(k)} \big\rangle \nonumber \\
	& \hspace{5cm} + \big\langle G^{(k)}_{\lambda_1} \big|\,\big(\nabla\chi^2\big) \!\cdot\! \nabla G_{\lambda_2}^{(k)} \big\rangle - \big\langle \nabla G^{(k)}_{\lambda_1} \big|\,\big(\nabla \chi^2\big) G_{\lambda_2}^{(k)} \big\rangle \bigg] \nonumber \\
%%%%%%%%%%%
	& \quad + 2 \sum_{k \in \{0,-1\}} q^{(k)} \left[ \lim_{r \to 0^{+}} \int_{\partial B_{r}(\mathbf{0})}\hspace{-0.5cm} d\Sigma_r\; e^{-i \mathbf{S}(\mathbf{0}) \cdot \mathbf{x}} \Big(i \chi \big(\mathbf{S} - \mathbf{S}(\mathbf{0})\big)\! \cdot\! \hat{\mathbf{r}} + \partial_r\chi \Big)\, \overline{\phi_{\lambda_2}}\, \big( G^{(k)}_{\lambda_2} - G^{(k)}_{\lambda_1} \big) \right] \nonumber \\
	& \quad + \sum_{k \in \{0,-1\}} q^{*}_{k} \bigg[ \lim_{r \to 0^{+}} \int_{\partial B_{r}(\mathbf{0})}\hspace{-0.5cm} d\Sigma_r\; e^{i \mathbf{S}(\mathbf{0}) \cdot \mathbf{x}} \Big( \big(i \chi \,\mathbf{S}(\mathbf{0}) \!\cdot\! \hat{\mathbf{r}} - \partial_{r}\chi \big)\, \overline{\big(G_{\lambda_2}^{(k)}\! - G_{\lambda_1}^{(k)}\big)}\, \phi_{\lambda_2} \nonumber \\
	& \hspace{3.5cm} + \chi \,\overline{\big(G_{\lambda_2}^{(k)}\! - G_{\lambda_1}^{(k)}\big)}\, \partial_r \phi_{\lambda_2} - \chi\, \overline{\partial_{r} \big(G_{\lambda_2}^{(k)}\! - G_{\lambda_1}^{(k)}\big)}\, \phi_{\lambda_2}\Big) \bigg] \nonumber \\
	& \quad + \sum_{k \in \{0,-1\}}  \big|q^{(k)}\big|^2\, \bigg[ -2i \,\lim_{r \to 0^{+}} \int_{\partial B_{r}(\mathbf{0})}\hspace{-0.5cm} d\Sigma_r\; \chi^2\big(\mathbf{S} - \mathbf{S}(\mathbf{0})\big) \!\cdot\! \hat{\mathbf{r}}\; \overline{\big(G_{\lambda_2}^{(k)}\! - G_{\lambda_1}^{(k)}\big)}\, G^{(k)}_{\lambda_1} \nonumber \\
	& \hspace{1.7cm} + \lim_{r \to 0^{+}} \int_{\partial B_{r}(\mathbf{0})}\hspace{-0.5cm} d\Sigma_r\;  \chi \partial_{r} \chi\, \bigg( \,\big|G^{(k)}_{\lambda_2}\! - G^{(k)}_{\lambda_1}\big|^2 - 2\,\mbox{Re} \Big(\overline{G^{(k)}_{\lambda_1}} \big(G_{\lambda_2}^{(k)}\! - G^{(k)}_{\lambda_1}\big)\Big) \bigg) \bigg] \,.
	\label{eq:proofQb}
\end{align}
}

\noindent
By comparison with \eqref{eq:QaSb}\eqref{eq:Xidef}, it appears that the terms from the second to the sixth line of \eqref{eq:proofQb} exactly reproduce $Q_{\alpha,S}^{(\beta)}\left[\phi_{\lambda_2}\! + e^{-i \mathbf{S}(\mathbf{0}) \cdot \mathbf{x}} \chi \mbox{$\sum_{k \in \{0,-1\}}$} q^{(k)} G_{\lambda_2}^{(k)}\right]$.
We now proceed to show that all other contributions vanish. On one side, consider the terms in the seventh and eighth lines of \eqref{eq:proofQb}. An additional integration by parts gives
{\small 
\begin{align*}
	& \big\langle G^{(k)}_{\lambda_1} \big|\,\big(\nabla\chi^2\big) \!\cdot\! \nabla G_{\lambda_2}^{(k)} \big\rangle - \big\langle \nabla G^{(k)}_{\lambda_1} \big|\,\big(\nabla \chi^2\big) G_{\lambda_2}^{(k)} \big\rangle \\
	& = \lim_{r \to 0^{+}} \left[ \int_{\partial B_{r}(\mathbf{0})}\hspace{-0.55cm} d\Sigma_r\, \chi^2 \left[\overline{\partial_r G^{(k)}_{\lambda_1}} G_{\lambda_2}^{(k)}\! - \overline{G^{(k)}_{\lambda_1}} \partial_r G_{\lambda_2}^{(k)}\right]+ \!\int_{\mathbb{R}^2 \setminus B_{r}(\mathbf{0})}\hspace{-0.6cm} d\mathbf{x}\; \chi^2 \left( \overline{\Delta G^{(k)}_{\lambda_1}} G_{\lambda_2}^{(k)} \! - \overline{G^{(k)}_{\lambda_1}} \Delta G_{\lambda_2}^{(k)} \right) \right].
\end{align*}
}

\noindent
From \eqref{eq:G01asy} we deduce, for $r \to 0^{+}$,
	\begin{equation*}
		\Big(\overline{\partial_r G^{(k)}_{\lambda_1}} G_{\lambda_2}^{(k)}\! - \overline{G^{(k)}_{\lambda_1}} \partial_r G_{\lambda_2}^{(k)}\Big)(r) = {\pi\, (\lambda_{2}^{2 |k+\alpha|}\! -\! \lambda_{1}^{2 |k+\alpha|}) \over 2 \sin(\pi \alpha)\, r} + \mathcal{O}\big(r^{1-2|k+\alpha|}\big)\,.
	\end{equation*}
Moreover, in view of \eqref{eq:defGEq} and \eqref{eq:G01exp}, an explicit computation gives
	\begin{equation*}
		\Delta G^{(k)}_{\lambda} \!
			= \big(\mathbf{A}^2 \!+ \lambda^2 + 2 \mathbf{A} \!\cdot\! (-i \nabla)\big) G^{(k)}_{\lambda}\!
			= \left(\mathbf{A}^2 \!+ \lambda^2 + {2\alpha k \over r}\right) G^{(k)}_{\lambda}, \quad
			\mbox{in\, $\mathbb{R}^2 \!\setminus\! \{\mathbf{0}\}$}\,.
	\end{equation*}
Recalling that $\chi \!=\! 1$ in an open neighborhood of $\mathbf{x} = \mathbf{0}$, see \eqref{eq:Hypchi}, we thus obtain
	\begin{align*}
		& {\pi^2 \over \sin(\pi \alpha)}\left(\lambda_1^{2 |k+\alpha|} \!-\! \lambda_2^{2 |k+\alpha|} \right) + (\lambda_2^2 - \lambda_1^2) \big\langle \chi G_{\lambda_1}^{(k)} \big|\, \chi  G_{\lambda_2}^{(k)} \big\rangle \\
		& \hspace{3.cm}  + \big\langle G^{(k)}_{\lambda_1} \big|\,\big(\nabla\chi^2\big) \!\cdot\! \nabla G_{\lambda_2}^{(k)} \big\rangle - \big\langle \nabla G^{(k)}_{\lambda_1} \big|\,\big(\nabla \chi^2\big) G_{\lambda_2}^{(k)} \big\rangle = 0\,.
	\end{align*}
On the other side, consider the boundary contributions in the last five lines of \eqref{eq:proofQb}. For $r$ small enough, the following holds true: $\partial_r \chi \!=\! 0$ on $\partial B_r(\mathbf{0})$, see \eqref{eq:Hypchi}; $|\mathbf{S} - \mathbf{S}(\mathbf{0})| \!\leqslant\! c\,r$, see \eqref{eq:SLip}; $|G^{(k)}_{\lambda_2}\! - G^{(k)}_{\lambda_1}| \!\leqslant\! c\,r^{|\alpha + k|}$ and $\big|\partial_r(G^{(k)}_{\lambda_2}\! - G^{(k)}_{\lambda_1})\big| \!\leqslant\! c\,r^{|\alpha + k|-1}$, see \eqref{eq:G01asy}. Recalling as well condition \eqref{eq:asypsiQF}, by Cauchy-Schwarz inequality we get:
{\small 
\begin{align*}
& \left|\int_{\partial B_{r}(\mathbf{0})}\hspace{-0.5cm} d\Sigma_r\; e^{-i \mathbf{S}(\mathbf{0}) \cdot \mathbf{x}} \Big(i \chi \big(\mathbf{S} - \mathbf{S}(\mathbf{0})\big)\! \cdot\! \hat{\mathbf{r}} + \partial_r\chi \Big)\, \overline{\phi_{\lambda_2}}\, \big( G^{(k)}_{\lambda_2}\! - G^{(k)}_{\lambda_1} \big)\right| \\
& \hspace{6cm} \leqslant c\,r^{2 + |\alpha + k|} \,\sqrt{\big\langle |\phi_{\lambda_2}|^2 \big\rangle} \;\xrightarrow{r \to 0^{+}}\, 0\,;
\end{align*}
\begin{align*}
& \bigg|\int_{\partial B_{r}(\mathbf{0})}\hspace{-0.5cm} d\Sigma_r\; e^{i \mathbf{S}(\mathbf{0}) \cdot \mathbf{x}} \Big( \big(i \chi \,\mathbf{S}(\mathbf{0}) \!\cdot\! \hat{\mathbf{r}} - \partial_{r}\chi \big)\, \overline{\big(G_{\lambda_2}^{(k)}\! - G_{\lambda_1}^{(k)}\big)}\, \phi_{\lambda_2} \\
	& \hspace{2.8cm} + \chi \,\overline{\big(G_{\lambda_2}^{(k)}\! - G_{\lambda_1}^{(k)}\big)}\, \partial_r \phi_{\lambda_2} - \chi\, \overline{\partial_{r} \big(G_{\lambda_2}^{(k)}\! - G_{\lambda_1}^{(k)}\big)}\, \phi_{\lambda_2}\Big)\bigg| \\
& \hspace{1.5cm} \leqslant c\,r^{|\alpha + k|} \left( r\, \sqrt{\big\langle |\phi_{\lambda_2}|^2 \big\rangle} + r\, \sqrt{\big\langle |\partial_{r} \phi_{\lambda_2}|^2 \big\rangle} + \sqrt{\big\langle |\phi_{\lambda_2}|^2 \big\rangle} \right) \;\xrightarrow{r \to 0^{+}}\, 0\,;
\end{align*}
\begin{align*}
\left|\int_{\partial B_{r}(\mathbf{0})}\hspace{-0.5cm} d\Sigma_r\; \chi^2\big(\mathbf{S} - \mathbf{S}(\mathbf{0})\big) \!\cdot\! \hat{\mathbf{r}}\; \overline{\big(G_{\lambda_2}^{(k)}\! - G_{\lambda_1}^{(k)}\big)}\, G^{(k)}_{\lambda_1} \right| \leqslant C r^{2} \;\xrightarrow{r \to 0^{+}}\, 0\,;
\end{align*}
\begin{align*}
\int_{\partial B_{r}(\mathbf{0})}\hspace{-0.5cm} d\Sigma_r\;  \chi \partial_{r} \chi\, \bigg( \,\big|G^{(k)}_{\lambda_2}\! - G^{(k)}_{\lambda_1}\big|^2 - 2\,\mbox{Re} \Big(\overline{G^{(k)}_{\lambda_1}} \big(G_{\lambda_2}^{(k)}\! - G^{(k)}_{\lambda_1}\big)\Big) \bigg) = 0\,.
\end{align*}
}

\noindent
Summing up, the previous results entail
	\begin{equation*}
		Q_{\alpha,S}^{(\beta)}\bigg[\phi_{\lambda_1}\! + e^{-i \mathbf{S}(\mathbf{0}) \cdot \mathbf{x}} \chi \!\!\sum_{k \in \{0,-1\}}\!\!\! q^{(k)} G_{\lambda_1}^{(k)}\bigg]
		= Q_{\alpha,S}^{(\beta)}\bigg[\phi_{\lambda_2}\! + e^{-i \mathbf{S}(\mathbf{0}) \cdot \mathbf{x}} \chi \!\!\sum_{k \in \{0,-1\}}\!\!\! q^{(k)} G_{\lambda_2}^{(k)}\bigg]\,,
	\end{equation*}
whence the thesis. By similar arguments it can be shown that the form does not depend on the choice of $\chi$, as long as hypotheses \eqref{eq:Hypchi} \eqref{eq:chirad} are fulfilled.

{\sl iii)} Closedness can be deduced by classical arguments \cite{CO18,Te90}, once lower boundedness has been proved. Therefore, the thesis follows as soon as we show that
	\begin{equation}\label{eq:Qlowb}
		Q_{\alpha,S}^{(\beta)}[\psi] + \lambda^2\,\|\psi\|_2^2 \geqslant 0\,, \qquad \mbox{for $\lambda > 0$ large enough}\,.
	\end{equation}
To this avail, by minor variations of the arguments described in \cite{CF20} (also recall \eqref{eq:GjL2}), we obtain the following for any $\varepsilon_1,\varepsilon_2,\varepsilon_3 \!\in\! (0,1)$ and suitable $c_1,c_2,c_3 \!>\! 0$:
{\small 
\begin{align*}
Q_{\alpha,S}^{(F)}[\phi_{\lambda}] \geqslant {1 \over 2}\, Q_{\alpha,S}^{(F)}[\phi_{\lambda}] + {1 - \varepsilon_1 \over 2}\,\left\| \mathbf{1}_{\chi} (- i \nabla + \mathbf{A}_{\alpha}) \phi_{\lambda} \right\|_2^2 - {1 - \varepsilon_1 \over 2\varepsilon_1}\, c_1\, \big\|\phi_{\lambda}\big\|_2^2\,;
\end{align*}
\begin{align*}
& \sum_{k \in \{0,-1\}} \mbox{Re} \left[ q^{(k)} \left\langle (- i \nabla + \mathbf{A}_{\alpha}) \phi_{\lambda} \left|\, e^{- i\, \mathbf{S}(\mathbf{0}) \cdot \mathbf{x}} \big(\big(\mathbf{S} - \mathbf{S}(\mathbf{0})\big)\chi - i \nabla\chi \big) G^{(k)}_{\lambda} \!\right.\right\rangle \right] \\
& \hspace{3cm} \geqslant - {\varepsilon_2 \over 8} \left\| \mathbf{1}_{\chi} (- i \nabla + \mathbf{A}_{\alpha}) \phi_{\lambda} \right\|_2^2 - {2c_2 \over \varepsilon_2}\! \sum_{k \in \{0,-1\}}\!\! |q^{(k)}|^2\, \lambda^{2|k+\alpha| - 2}\, ;
\end{align*}
\begin{align*}
& \sum_{k \in \{0,-1\}}\!\!\! \mbox{Re} \left[ q^{(k)}\! \left\langle \phi_{\lambda} \left|\, e^{-i\, \mathbf{S}(\mathbf{0}) \cdot \mathbf{x}} \big[\big(\mathbf{S} - \mathbf{S}(\mathbf{0})\big)^2 \chi + 2\mathbf{S}(\mathbf{0}) \!\cdot\! \big(\big(\mathbf{S} - \mathbf{S}(\mathbf{0})\big)\chi - i \nabla\chi \big) + \Delta \chi \big] G^{(k)}_{\lambda}\! \right.\right\rangle  \right] \nonumber\\
& \hspace{5.5cm} \geqslant - {\varepsilon_3 \over 2}\, \|\phi_{\lambda}\|_{2}^{2} - {c_3 \over \varepsilon_3} \! \sum_{k \in \{0,-1\}}\!\! |q^{(k)}|^2\, \lambda^{2|k+\alpha| - 2}\, .
\end{align*}}

\noindent
Building on the basic inequality $\big|\nabla \big(\chi\, G^{(k)}_{\lambda}\big)\big| \!\leqslant\! {c \over |\mathbf{x}|}\, \big(\chi\, G^{(k)}_{\lambda}\big)$ and \eqref{eq:GjL2}\eqref{eq:SLip}, we further deduce
$ |\Xi_{kk'}(\lambda)| \!\leqslant\! c\,\|G^{(k)}_{\lambda}\|_{2}\, \|G^{(k')}_{\lambda}\|_{2} \!\leqslant\! c\, \lambda^{|k+\alpha|+|k'+\alpha|-2}$.
This allows us to infer that, for some suitable $c_4 \!>\! 0$,
	\begin{align*}
		& \sum_{k,k' \in \{0,-1\}}\!\! \overline{q^{(k)}}\, q^{(k')}\! \left[ \beta_{kk'} + {\pi^2 \over \sin(\pi \alpha)}\, \lambda^{2 |k+\alpha|}\, \delta_{kk'} + \Xi_{kk'}(\lambda) \right]  \\
		& \geqslant \left[{\pi^2 \over \sin(\pi \alpha)} \min_{k \in \{0,-1\}}\! \big(\lambda^{2 |k+\alpha|}\big) - \max_{k,k' \in \{0,-1\}}\! \Big(\, \big| \beta_{kk'}\big| + \big|\Xi_{kk'}(\lambda)\big|\, \Big) \right] \sum_{k \in \{0,-1\}} \big|q^{(k)}\big|^2 \\
		& \geqslant c_4\,\Big(\min\!\big\{\lambda^{2\alpha},\lambda^{2(1-\alpha)}\big\} -1 - \max\!\big\{\lambda^{-2\alpha},\lambda^{-2(1-\alpha)}\big\}\, \Big) \sum_{k \in \{0,-1\}} \big|q^{(k)}\big|^2 \,.
\end{align*}
Summing up, we have
{\small 
\begin{align*}
	& Q_{\alpha,S}^{(\beta)}[\psi] + \lambda^2\,\|\psi\|_2^{2} \\
	& \geqslant {1 \over 2}\, Q_{\alpha,S}^{(F)}[\phi_{\lambda}]
		+ {1 \!-\! \varepsilon_1 \!-\! \varepsilon_2 \over 2} \left\| \mathbf{1}_{\chi} (- i \nabla + \mathbf{A}_{\alpha}) \phi_{\lambda} \right\|_2^2
		+ \left(\lambda^2 - {1 - \varepsilon_1 \over 2\varepsilon_1}\, c_1 - \varepsilon_3 \right) \|\phi_{\lambda}\|_{2}^{2}  \\
	& \hspace{0.4cm} +  \left[c_4 \min\!\big\{\lambda^{2\alpha}\!,\lambda^{2(1-\alpha)}\big\} - c_4 - \left({8c_2 \over \varepsilon_2} \!+\! {2c_3 \over \varepsilon_3} \!+\! c_4 \right) \max\!\big\{\lambda^{-2\alpha}\!,\lambda^{-2(1-\alpha)} \big\}  \right] \sum_{k \in \{0,-1\}}\!\! \big|q^{(k)}\big|^2\,.
\end{align*}}

\noindent
Upon fixing $\varepsilon_1,\varepsilon_2,\varepsilon_3\!\in\!(0,1)$ appropriately and $\lambda \!>\! 0$ large enough, the above relation suffices to infer \eqref{eq:Qlowb}, whence the thesis.
\end{proof}

\begin{proof}[Corollary \ref{cor:HasB}]
For any Hermitian matrix $\beta$, we derive the self-adjoint operator $H_{\alpha,S}^{(\beta)}$ associated to the quadratic form $Q_{\alpha,S}^{(\beta)}$ by standard methods.
To begin with, for any pair $\psi_\ell = \phi_{\ell,\lambda} + e^{-i\, \mathbf{S}(\mathbf{0}) \cdot \mathbf{x}}\,\chi \sum_{k \in \{0,-1\}}  q_{\ell}^{(k)} G^{(k)}_{\lambda}$, $\ell \!\in \!\{1,2\}$, belonging to the form domain $D\big[Q_{\alpha,S}^{(\beta)}\big]$, consider the sesquilinear form defined by polarization
{\small \begin{align}
	& Q_{\alpha,S}^{(\beta)}[\psi_1,\psi_2] = Q_{\alpha,S}^{(F)}[\phi_{1,\lambda},\phi_{2,\lambda}] - \lambda^2 \langle \psi_1|\psi_2\rangle + \lambda^2 \langle \phi_{1,\lambda} | \phi_{2,\lambda} \rangle \nonumber \\
	& \hspace{0.3cm} +\! \sum_{k \in \{0,-1\}} \!\!\overline{q_1^{(k)}} \left[ 2 \left\langle \left. e^{- i\, \mathbf{S}(\mathbf{0}) \cdot \mathbf{x}} \big(\big(\mathbf{S} - \mathbf{S}(\mathbf{0})\big)\chi - i \nabla\chi \big) G^{(k)}_{\lambda} \right|\, (- i \nabla + \mathbf{A}_{\alpha}) \phi_{2,\lambda} \right\rangle \right.\nonumber \\
	& \hspace{1.5cm} \left. + \, \left\langle\left. e^{-i\, \mathbf{S}(\mathbf{0}) \cdot \mathbf{x}} \big[\big(\mathbf{S} - \mathbf{S}(\mathbf{0})\big)^2 \chi + 2\mathbf{S}(\mathbf{0}) \!\cdot\! \big(\big(\mathbf{S} - \mathbf{S}(\mathbf{0})\big)\chi - i \nabla\chi \big) + \Delta \chi \big] G^{(k)}_{\lambda} \right|\, \phi_{2,\lambda}\! \right\rangle \right]	\nonumber \\
	& \hspace{0.3cm} +\! \sum_{k \in \{0,-1\}} \!\!q_2^{(k)} \left[ 2 \left\langle (- i \nabla + \mathbf{A}_{\alpha}) \phi_{1,\lambda} \left|\, e^{- i\, \mathbf{S}(\mathbf{0}) \cdot \mathbf{x}} \big(\big(\mathbf{S} - \mathbf{S}(\mathbf{0})\big)\chi - i \nabla\chi \big) G^{(k)}_{\lambda} \!\right.\right\rangle \right.\nonumber \\
	& \hspace{1.5cm} \left. + \, \left\langle \phi_{1,\lambda} \left|\, e^{-i\, \mathbf{S}(\mathbf{0}) \cdot \mathbf{x}} \big[\big(\mathbf{S} - \mathbf{S}(\mathbf{0})\big)^2 \chi + 2\mathbf{S}(\mathbf{0}) \!\cdot\! \big(\big(\mathbf{S} - \mathbf{S}(\mathbf{0})\big)\chi - i \nabla\chi \big) + \Delta \chi \big] G^{(k)}_{\lambda}\! \right.\right\rangle \right] \nonumber \\
	& \hspace{0.3cm} + \sum_{k,k' \in \{0,-1\}} \overline{q_1^{(k)}} q_2^{(k')} \left[ \beta_{kk'} + {\pi^2 \over \sin(\pi \alpha)}\, \lambda^{2 |k+\alpha|}\, \delta_{kk'} + \Xi_{kk'}(\lambda) \right] . \label{eq:QaSbses}
\end{align}}

\noindent
Here $Q_{\alpha,S}^{(F)}[\phi_{1,\lambda},\phi_{2,\lambda}]$ is the sesquilinear form associated to the Friedrichs quadratic form, namely, $Q_{\alpha,S}^{(F)}[\phi_{1,\lambda},\phi_{2,\lambda}] := \int_{\mathbb{R}^2}d\mathbf{x}\, \overline{(-i\nabla\! + \mathbf{A}_{\alpha}\! + \mathbf{S})\phi_{1,\lambda}} \cdot (-i\nabla\! + \mathbf{A}_{\alpha}\! + \mathbf{S})\phi_{2,\lambda}\,$.\\
Now assume $q_1^{(0)} \!= q_1^{(-1)} \!= 0$, so that $\psi_1 \!=\! \phi_{1,\lambda}$. Integrating by parts and checking that boundary contributions vanish by means of arguments similar to those outlined in the proof of Theorem \ref{thm:QasB}, item \emph{ii)}, the sesquilinear form \eqref{eq:QaSbses} reduces to
	\begin{align*}
		& Q_{\alpha,S}^{(\beta)}[\phi_1,\psi_2] = \left\langle \phi_{1,\lambda} \left|\,H_{\alpha,S} \phi_{2,\lambda} \right.\right\rangle  \nonumber \\
			& \hspace{0.cm} +\! \sum_{k \in \{0,-1\}} \!\!q_2^{(k)} \left[ 
		2 \left\langle \phi_{1,\lambda} \left|\,e^{- i\, \mathbf{S}(\mathbf{0}) \cdot \mathbf{x}}  \big(\big(\mathbf{S} - \mathbf{S}(\mathbf{0})\big)\chi - i \nabla\chi \big) \cdot \big(- i \nabla + \mathbf{A}_{\alpha}\big) G^{(k)}_{\lambda} \!\right.\right\rangle \right.\nonumber \\
			& \hspace{0.5cm} \left. + \, \left\langle \phi_{1,\lambda} \left|\, e^{-i\, \mathbf{S}(\mathbf{0}) \cdot \mathbf{x}} \Big(\big(\mathbf{S} - \mathbf{S}(\mathbf{0})\big)^2 \chi - \lambda^2 \chi + 2\big(\mathbf{S} - \mathbf{S}(\mathbf{0})\big)\!\cdot\!(- i \nabla\chi) - \Delta \chi \Big) G^{(k)}_{\lambda}\! \right.\right\rangle \right] .
	\end{align*}
Considerations analogous to those reported in the proof of Theorem \ref{thm:QasB}, item \emph{i)}, ensure that all pairings in the second and third lines of the above identity are well-defined inner products in $L^2(\mathbb{R}^2)$. So, to fulfill the condition $Q_{\alpha,S}^{(\beta)}[\phi_1,\psi_2]\! =\! \langle \phi_1 | w\rangle$ for some $w = H_{\alpha,S}^{(\beta)}\, \psi_2 \!\in\! L^2(\mathbb{R}^2)$, we must require $H_{\alpha,S}\, \phi_{2,\lambda} \!\in\! L^2(\mathbb{R}^2)$ (\emph{cf.} \eqref{eq:DHaSF} and the condition in the second line of \eqref{eq:DHaSbe}), as well as (\emph{cf.} \eqref{eq:HaSbe})
	\begin{align}
		& w = H_{\alpha,S}\, \phi_{2,\lambda} +\!\! \sum_{k \in \{0,-1\}} \!\!\!q_2^{(k)} e^{- i\, \mathbf{S}(\mathbf{0}) \cdot \mathbf{x}} \!\left[  2 \Big(\big(\mathbf{S} - \mathbf{S}(\mathbf{0})\big)\chi - i \nabla\chi \Big) \!\cdot\! \big(\!- i \nabla + \mathbf{A}_{\alpha}\big) G^{(k)}_{\lambda} \right.\nonumber \\
			& \hspace{2.cm} \left. + \, \Big(\big(\mathbf{S} - \mathbf{S}(\mathbf{0})\big)^2 \chi - \lambda^2 \chi + 2\big(\mathbf{S} - \mathbf{S}(\mathbf{0})\big)\!\cdot\!(- i \nabla\chi) - \Delta \chi \Big) G^{(k)}_{\lambda}\right] . \label{eq:wdef}
	\end{align}
In view of the previous results, the sesquilinear form \eqref{eq:QaSbses} can be re-written as
{\small \begin{align*}
	& Q_{\alpha,S}^{(\beta)}[\psi_1,\psi_2] = Q_{\alpha,S}^{(\beta)}[\phi_{1,\lambda},\psi_2] \\
	& \hspace{0.3cm} +\! \sum_{k \in \{0,-1\}} \!\!\overline{q_1^{(k)}} \left[ 2 \left\langle\left. e^{- i\, \mathbf{S}(\mathbf{0}) \cdot \mathbf{x}} \big(\big(\mathbf{S} - \mathbf{S}(\mathbf{0})\big)\chi - i \nabla\chi \big) G^{(k)}_{\lambda} \right|\, (- i \nabla + \mathbf{A}_{\alpha}) \phi_{2,\lambda} \right\rangle \right.\nonumber \\
	& \hspace{0.9cm} \left. + \, \left\langle\left. e^{-i\, \mathbf{S}(\mathbf{0}) \cdot \mathbf{x}} \big[\big(\mathbf{S} - \mathbf{S}(\mathbf{0})\big)^2 \chi - \lambda^2 \chi + 2\mathbf{S}(\mathbf{0}) \!\cdot\! \big(\big(\mathbf{S} - \mathbf{S}(\mathbf{0})\big)\chi - i \nabla\chi \big) + \Delta \chi \big] G^{(k)}_{\lambda} \right|\, \phi_{2,\lambda}\! \right\rangle \right]	\nonumber \\
	& \hspace{0.3cm} + \sum_{k,k' \in \{0,-1\}} \overline{q_1^{(k)}} q_2^{(k')} \left[ \beta_{kk'} + {\pi^2 \over \sin(\pi \alpha)}\, \lambda^{2 |k+\alpha|}\, \delta_{kk'} + \Xi_{kk'}(\lambda) - \lambda^2 \left\langle \chi\, G^{(k)}_{\lambda} \left|\, \chi \, G^{(k')}_{\lambda} \right.\right\rangle \right] .
\end{align*}}

\noindent
Building on this and recalling the definition \eqref{eq:Xidef} of $\Xi_{kk'}(\lambda)$, by simple (though lengthy) computations we deduce that the position $Q_{\alpha,S}^{(\beta)}[\psi_1,\psi_2] \!=\! \langle \psi_1|w \rangle$, with $w$ as in \eqref{eq:wdef}, can be satisfied for generic $q_1^{(0)}\!,q_1^{(-1)}$ only if, for $k\!\in\!\{0,-1\}$,
	\begin{align}\label{eq:proofbc}
		& \left\langle G^{(k)}_{\lambda} \left|\, \big[(-i \nabla + \mathbf{A}_{\alpha})^2 + \lambda^2 \big] \big( e^{i\, \mathbf{S}(\mathbf{0}) \cdot \mathbf{x}}\,\chi\, \phi_{2,\lambda} \big) \right. \right\rangle \nonumber \\
		& = \sum_{k' \in \{0,-1\}}\!\! q_2^{(k')}\, \bigg[ \,\beta_{kk'} + {\pi^2 \over \sin(\pi \alpha)}\, \lambda^{2 |k+\alpha|}\,\delta_{kk'} \; +  \nonumber \\
			& \hspace{1.2cm} + \bigg(\, \left\langle G^{(k)}_{\lambda} \left|\, \nabla \!\cdot\! \big(\chi \nabla \chi\big)\, G^{(k)}_{\lambda}\right.\right\rangle + 2 \left\langle G^{(k)}_{\lambda} \left|\, (\chi \nabla\chi) \!\cdot\! \nabla G^{(k)}_{\lambda} \right.\right\rangle \,\bigg)\, \delta_{kk'} \bigg]\, .
	\end{align}
To derive the above identity we used in particular the identity \eqref{eq:GkGkrad} and the fact that $\mathbf{A}_{\alpha} \cdot \nabla\chi = 0$, both descending from \eqref{eq:chirad}. On one side, recalling the explicit expression \eqref{eq:G01exp} for $G^{(k)}_{\lambda}$ and that $\chi$ is radial, we get $2\,\overline{G^{(k)}_{\lambda} } (\chi \nabla\chi) \cdot \nabla G^{(k)}_{\lambda} \!= (\chi \nabla\chi) \cdot \nabla \big|G^{(k)}_{\lambda}\big|^2$; then, integrating by parts and keeping in mind that $\chi \!\equiv\! 1$ near the origin, we obtain
	\begin{align}
		& \left\langle G^{(k)}_{\lambda} \left|\, \nabla \!\cdot\! \big(\chi \nabla \chi\big)\, G^{(k)}_{\lambda}\right.\right\rangle + 2 \left\langle G^{(k)}_{\lambda} \left|\, (\chi \nabla\chi) \!\cdot\! \nabla G^{(k)}_{\lambda} \right.\right\rangle \nonumber \\
		& \hspace{4.cm} = - \lim_{r \to 0^{+}} \int_{\partial B_{r}(\mathbf{0})}\hspace{-0.5cm} d\Sigma_r\, \big(\chi \partial_r \chi\big)\, \big|G^{(k)}_{\lambda}\big|^2 = 0\,. \label{eq:Gnab0}
	\end{align}
On the other side, integrating by parts twice and using the basic identity \eqref{eq:defGEq}, we get
	\begin{align}
		& \left\langle G^{(k)}_{\lambda} \left|\, \big[(-i \nabla + \mathbf{A}_{\alpha})^2 + \lambda^2 \big] \big( e^{i\, \mathbf{S}(\mathbf{0}) \cdot \mathbf{x}}\,\chi\, \phi_{2,\lambda} \big) \right. \right\rangle \nonumber \\
		& = \lim_{r \to 0^{+}} \int_{\partial B_{r}(\mathbf{0})}\hspace{-0.5cm} d\Sigma_r \left[\overline{G^{(k)}_{\lambda}}\, \partial_{r} \big( e^{i\, \mathbf{S}(\mathbf{0}) \cdot \mathbf{x}}\,\chi\, \phi_{2,\lambda} \big) -\overline{\partial_r G^{(k)}_{\lambda}}\,  \big( e^{i\, \mathbf{S}(\mathbf{0}) \cdot \mathbf{x}}\,\chi\, \phi_{2,\lambda} \big) \right] \nonumber  \\
			& \hspace{3cm} + \lim_{r \to 0^{+}} \int_{\mathbb{R}^2 \setminus B_{r}(\mathbf{0})}\hspace{-0.5cm} d\Sigma_r\, \overline{\big[ (-i \nabla + \mathbf{A}_{\alpha})^2 + \lambda^2 \big] G^{(k)}_{\lambda}}\, \big( e^{i\, \mathbf{S}(\mathbf{0}) \cdot \mathbf{x}}\,\chi\, \phi_{2,\lambda} \big) \nonumber  \\
		& = \lim_{r \to 0^{+}} \int_{\partial B_{r}(\mathbf{0})}\hspace{-0.5cm} d\Sigma_r \left[\overline{G^{(k)}_{\lambda}}\, \partial_{r} \big( e^{i\, \mathbf{S}(\mathbf{0}) \cdot \mathbf{x}}\,\chi\, \phi_{2,\lambda} \big) -\overline{\partial_r G^{(k)}_{\lambda}}\, \big( e^{i\, \mathbf{S}(\mathbf{0}) \cdot \mathbf{x}}\,\chi\, \phi_{2,\lambda} \big) \right] \nonumber \\
		& = {\Gamma\big( |k+\alpha| \big) \over 2^{1-|k+\alpha|}}\, \lim_{r \to 0^{+}} {1 \over r^{|k+\alpha|}} \int_{\partial B_{r}(\mathbf{0})}\hspace{-0.3cm} d\Sigma_r\; e^{- i k \theta} \left[\partial_{r}\phi_{2,\lambda} 
		+ {|k+\alpha| \over r}\, \phi_{2,\lambda} \right], \label{eq:pbc}
	\end{align}
where the last identity follows from the asymptotic relations \eqref{eq:asypsiQF} \eqref{eq:G01asy}, by arguments analogous to those mentioned in the proof of Theorem \ref{thm:QasB}. Notably, only the leading order term in \eqref{eq:G01asy} plays a role here.
Summing up, from \eqref{eq:proofbc}\eqref{eq:Gnab0} and \eqref{eq:pbc} we infer
	\begin{align*}
		& \sum_{k' \in \{0,-1\}}\! q_2^{(k')}\, \bigg( \,\beta_{kk'} + {\pi^2 \over \sin(\pi \alpha)}\, \lambda^{2 |k+\alpha|}\,\delta_{kk'} \bigg) \\
		& \hspace{1.5cm} = {\Gamma\big( |k+\alpha| \big) \over 2^{1-|k+\alpha|}}\, \lim_{r \to 0^{+}} {1 \over r^{|k+\alpha|}} \int_{\partial B_{r}(\mathbf{0})}\hspace{-0.3cm} d\Sigma_r\; e^{- i k \theta} \left[\partial_{r}\phi_{2,\lambda} + {|k+\alpha| \over r}\, \phi_{2,\lambda} \right] \,,
	\end{align*}
which proves the boundary condition in \eqref{eq:DHaSbe}, thus completing the characterization of $D\big(H_{\alpha,S}^{(\beta)}\big)$.
\vspace{0.05cm}

The fact that the family $H_{\alpha,S}^{(\beta)}$, $\beta$ any $2\!\times\! 2$ Hermitian matrix, exhausts all self-adjoint extensions of $H_{\alpha,S}\! \upharpoonright\! C^{\infty}_c(\mathbb{R}^2 \setminus \{\mathbf{0}\})$ in $L^2(\mathbb{R}^2)$ if $\mathbf{S} \!\in\! L^{\infty}(\mathbb{R}^2)$ can be deduced by exactly the same arguments reported in \cite[Proof of Corollary 1.10]{CF20}.\,
\end{proof}

Let us finally proceed to present the proof of Theorem \ref{thm:Gconv}, keeping in mind that it relies on the hypothesis \eqref{eq:SLipinf}. The latter implies that $\mathbf{S}$ is uniformly bounded on the whole space $\mathbb{R}^2$ and Lipschitz continuous at the origin.

\begin{proof}[Theorem \ref{thm:Gconv}]
The derivation of both the lower and upper bound inequalities relies on the following algebraic identity, which can be easily deduced using the gauge transformation $\psi \mapsto e^{-i \mathbf{S}(\mathbf{0}) \cdot \mathbf{x}}\,\psi$ and an elementary telescopic argument:
	\begin{align}
		Q_{\alpha_n,S}^{(F)}[\psi_{\alpha_n}] - Q_{0,S}^{(F)}[\psi_0]
		& = \left\| \mathbf{A}_{\alpha_n} \psi_{\alpha_n} \right\|_{2}^{2} + 2\, \mbox{Re}\! \left[ \left\langle \big(- i \nabla + \mathbf{S} - \mathbf{S}(\mathbf{0})\big) \psi_{\alpha_{n}} \left| \mathbf{A}_{\alpha_n} \psi_{\alpha_{n}} \right. \right\rangle \right] \nonumber \\
				& \quad  + \left\|\big(\!- i \nabla + \mathbf{S} - \mathbf{S}(\mathbf{0})\big) \psi_{\alpha_{n}} \right\|_{2}^2\! - \left\|\big(\!- i \nabla + \mathbf{S} - \mathbf{S}(\mathbf{0})\big) \psi_{0} \right\|_{2}^2 . \nonumber \\
		\label{eq:teles}
	\end{align}

{\sl i) Lower bound inequality.} First of all, on account of the hypothesis $\mathbf{S} \!\in\! L^{\infty}(\mathbb{R}^2)$, from \cite[Eq. (2.7)]{CF20} we deduce that
	\begin{equation}
		Q_{\alpha_n,S}^{(F)}[\psi_{\alpha_n}] + \gamma\,\|\psi_{\alpha_n}\|_2^2 \geqslant C_{\gamma}\! \left(\,\left\| \nabla \psi_{\alpha_n} \right\|_{2}^{2} + \left\| \mathbf{A}_{\alpha_n} \psi_{\alpha_n} \right\|_{2}^{2} \,\right), \label{eq:pr1}
	\end{equation}
for any $\gamma\!>\!0$ large enough and some suitable $C_{\gamma} \!>\! 0$. With obvious understandings, the above inequality is in fact valid for all $\psi_{\alpha_n}\!\!\in\!L^2(\mathbb{R}^2)$.

For any convergent sequence $\psi_{\alpha_n} \!\!\to\! \psi_0 \!\in\! L^2(\mathbb{R}^2) \!\setminus\! H^1(\mathbb{R}^2)$, the thesis can be derived by \emph{reductio ad absurdum}. In this case, the condition \eqref{eq:lowerb} reads (\emph{cf.} \eqref{eq:QaSext})
	\begin{equation*}
		+\infty = Q_{0,S}^{(F)}[\psi_0] \leqslant \liminf_{n \to +\infty} Q_{\alpha_n,S}^{(F)}[\psi_{\alpha_n}]
		\leqslant \limsup_{n \to +\infty} Q_{\alpha_n,S}^{(F)}[\psi_{\alpha_n}]\,.
	\end{equation*}
By contradiction, assume there exists some sequence $\psi_{\alpha_n} \!\!\to\! \psi_0\!\in\! L^2(\mathbb{R}^2) \!\setminus\! H^1(\mathbb{R}^2)$ such that $\lim_{n \to +\infty} Q_{\alpha_n,S}^{(F)}[\psi_{\alpha_n}] \!\leqslant \!C_S \!<\! + \infty$. Then, from \eqref{eq:pr1} it follows that the said sequence is uniformly bounded in $H^1(\mathbb{R}^2)$. By Banach-Alaoglu theorem, this implies in turn that $\psi_{\alpha_n} \!\!\rightharpoonup\! \varphi \!\in\!H^1(\mathbb{R}^2)$ (weak convergence, up to extraction of a subsequence). This contradicts the hypothesis $\psi_{\alpha_n} \!\!\to\! \psi_0 \!\in\! L^2(\mathbb{R}^2) \!\setminus\! H^1(\mathbb{R}^2)$, since uniqueness of the limit implies $\psi_0 = \varphi$.

Next, consider any convergent sequence in $L^2(\mathbb{R}^2)$ fulfilling $\psi_{\alpha_n} \!\!\to\! \psi_0 \!\in\! H^1(\mathbb{R}^2)$. The thesis \eqref{eq:lowerb} follows trivially if $Q_{\alpha_n,S}^{(F)}[\psi_{\alpha_n}] \!>\! Q_{0,S}^{(F)}[\psi_{0}]$ for almost all $n \!\in\! \mathbb{N}$. On the contrary, let us assume that $Q_{\alpha_n,S}^{(F)}[\psi_{\alpha_n}] \!\leqslant\! Q_{0,S}^{(F)}[\psi_{0}]$ for almost all $n \!\in\! \mathbb{N}$. Since $Q_{0,S}^{(F)}[\psi_{0}] \!<\! +\infty$ for $\psi_{0}\!\in\!H^1(\mathbb{R}^2)$, by arguments similar to those described before we deduce the existence of a uniformly bounded subsequence $\{\psi_{\tilde{\alpha}_n}\}_{n \,\in\, \mathbb{N}}$ in $H^1(\mathbb{R}^2)$, converging weakly to $\psi_0$. Taking this into account, let us now refer to \eqref{eq:teles}. On one side, notice that $\|(- i \nabla + \mathbf{S} - \mathbf{S}(\mathbf{0}))\psi\|_2^2 + \|\psi\|_2^2$ defines an equivalent norm in $H^1(\mathbb{R}^2)$ for $\mathbf{S}\!\in\!L^{\infty}(\mathbb{R}^2)$. Then, keeping in mind that $\psi_{\tilde{\alpha}_n} \!\!\to\! \psi_0$ in the strong $L^2$-topology, by lower semicontinuity of the norm in $H^1(\mathbb{R}^2)$ we infer
	\begin{align*}
		& \liminf_{n \to +\infty} \left\|\big(- i \nabla + \mathbf{S} - \mathbf{S}(\mathbf{0})\big) \psi_{\tilde{\alpha}_{n}} \right\|^2_2 \\
		& = \liminf_{n \to +\infty} \left(\, \left\|\big(- i \nabla + \mathbf{S} - \mathbf{S}(\mathbf{0})\big) \psi_{\tilde{\alpha}_{n}} \right\|^2_2 + \left\|\psi_{\tilde{\alpha}_{n}} \right\|^2_2 \right) - \lim_{n \to +\infty} \left\|\psi_{\tilde{\alpha}_{n}} \right\|^2_2 \\
		& \geqslant \left(\left\|\big(- i \nabla + \mathbf{S} - \mathbf{S}(\mathbf{0})\big) \psi_{0} \right\|^2_2 + \left\|\psi_{0} \right\|^2_2 \right) - \left\|\psi_{0} \right\|^2_2
		= \left\|\big(- i \nabla + \mathbf{S} - \mathbf{S}(\mathbf{0})\big) \psi_{0} \right\|^2_2 \,.
	\end{align*}
On the other side, using the angular harmonics decomposition
	\begin{equation*}
		\psi_{\tilde{\alpha}_n}(r,\theta) = \sum_{k \in \mathbb{Z}} \psi_{\tilde{\alpha}_n}^{(k)}(r)\,{e^{i k \theta} \over \sqrt{2\pi}}\;,
	\end{equation*}
by a direct computation we infer
	\begin{align*}
		& \left| \left\langle (- i \nabla) \psi_{\tilde{\alpha}_{n}} \left|\, \mathbf{A}_{\tilde{\alpha}_n} \psi_{\tilde{\alpha}_{n}} \right. \right\rangle \right|
			= \left|\, \sum_{k \in \mathbb{Z}} \int_{0}^{+\infty}\!\!\!dr\;{\tilde{\alpha}_n\, k \over r}\,\big|\psi_{\tilde{\alpha}_n}^{(k)}(r)\big|^2 \,\right| \\
		& \hspace{4cm} \leqslant \tilde{\alpha}_n \sum_{k \in \mathbb{Z}} \int_{0}^{+\infty}\!\!\!dr\;{k^2 \over r}\,\big|\psi_{\tilde{\alpha}_n}^{(k)}(r)\big|^2
			\leqslant \tilde{\alpha}_n \left\|\psi_{\tilde{\alpha}_{n}}\right\|_{H^1}^2 .
	\end{align*}
At the same time, exploiting the Lipschitz continuity of $\mathbf{S}$ at $\mathbf{x}=\mathbf{0}$, we get
	\begin{align*}
		\left|\, \left\langle \big(\mathbf{S} - \mathbf{S}(\mathbf{0})\big) \psi_{\tilde{\alpha}_{n}} \left|\, \mathbf{A}_{\tilde{\alpha}_{n}} \psi_{\tilde{\alpha}_{n}} \right. \right\rangle \right| 
		\leqslant \left\|\big(\mathbf{S} - \mathbf{S}(\mathbf{0})\big) \!\cdot\! \mathbf{A}_{\tilde{\alpha}_{n}}\right\|_{\infty}\, \left\| \psi_{\tilde{\alpha}_{n}}\right\|_{2}^2
		\leqslant \tilde{\alpha}_{n}\,c \left\| \psi_{\tilde{\alpha}_{n}}\right\|_{2}^2 .
	\end{align*}
Discarding the positive term $\|\mathbf{A}_{\alpha_n} \psi_{\alpha_n}\|_{2}^{2}$ and recalling that $\{\psi_{\tilde{\alpha}_{n}}\}_{n\,\in\,\mathbb{N}}$ is uniformly bounded in $H^1(\mathbb{R}^2)$, from the above arguments and \eqref{eq:teles} we deduce
	\begin{align*}
		\liminf_{n\to + \infty} Q_{\alpha_n,S}^{(F)}[\psi_{\alpha_n}] - Q_{0,S}^{(F)}[\psi_0]
		& \geqslant -\,2 \limsup_{n\to + \infty}  \left|\left\langle \big(\!- i \nabla + \mathbf{S} - \mathbf{S}(\mathbf{0})\big) \psi_{\alpha_{n}} \left| \mathbf{A}_{\alpha_n} \psi_{\alpha_{n}} \right. \right\rangle \right| \\
		& \geqslant -\,C \limsup_{n \to + \infty} \left( \tilde{\alpha}_{n}\, \left\| \psi_{\tilde{\alpha}_{n}}\right\|_{H^1}^2 \right) = 0\,, \nonumber
	\end{align*}
which proves the lower bound inequality \eqref{eq:lowerb}.

{\sl ii) Upper bound inequality.} For $\psi_0 \!\in\! L^2(\mathbb{R}^2) \!\setminus\! H^1(\mathbb{R}^2)$ the thesis \eqref{eq:upperb} is trivial, since $Q_{0,S}^{(F)}[\psi_0] \!=\! +\infty$ by \eqref{eq:Q0Sext}. Let us henceforth assume $\psi_0 \!\in\! H^1(\mathbb{R}^2)$. For any given family $\{\alpha_n\}_{n \,\in\, \mathbb{N}} \!\subset\! (0,1)$ such that $\alpha_n \!\to\! 0$ as $n \!\to\! +\infty$, we consider the sequence of approximants
	\begin{equation*}
		\psi_{\alpha_n} \!:= \eta_{\alpha_n}\,\psi_0 \in L^2(\mathbb{R}^2)\,,
	\end{equation*}
where $\eta_{\alpha_n}(\mathbf{x}) \!\equiv\! \eta_{\alpha_n}\big(|\mathbf{x}|\big) : [0,+\infty) \to [0,1]$ is a monotone increasing, smooth radial function with downward concavity fulfilling
	\begin{equation}\label{eq:eta01}
		\eta_{\alpha_n}(\mathbf{x}) = \left\{\, \begin{array}{ll}
			\displaystyle{\left(\,|\mathbf{x}|/\sqrt{\alpha_n}\,\right)^{\alpha_n}} & \quad \mbox{for\, $\mathbf{x} \!\in\! B_{\sqrt{\alpha_n}}(\mathbf{0})$}\,, \vspace{0.1cm}\\
			\displaystyle{1} & \quad \mbox{for\, $\mathbf{x} \!\in\! \mathbb{R}^2 \!\setminus\! B_{2\sqrt{\alpha_n}}(\mathbf{0})$}\,.
		\end{array}\right.
	\end{equation}
By monotone convergence, we readily infer
	\begin{equation*}
		\left\|\psi_{\alpha_n} - \psi_{0}\right\|_2^{2} \,= \int_{\mathbb{R}^2}\! d\mathbf{x}\;\left|\eta_{\alpha_n} - 1 \right|^2 \left|\psi_{0}\right|^2 \;\xrightarrow{n \to +\infty}\; 0\,,
	\end{equation*}
proving the required strong convergence $\psi_{\alpha_n} \!\to \psi_{0}$ in $L^2(\mathbb{R}^2)$.\\
In the sequel we proceed to deduce the upper bound \eqref{eq:upperb}, using the telescopic identity \eqref{eq:teles} to derive the stronger condition
	\begin{equation}\label{eq:proofx}
		\lim_{n \to +\infty} \left|Q_{\alpha_n,S}^{(F)}[\psi_{\alpha_n}] - Q_{0,S}^{(F)}[\psi_0] \right| = 0\,.
	\end{equation}
To this purpose, let us first consider the expression $\left\| \mathbf{A}_{\alpha_n} \psi_{\alpha_n} \right\|_{2}^2$ in \eqref{eq:teles} and refer to the decomposition
	\begin{equation*}
		\left\| \mathbf{A}_{\alpha_n} \psi_{\alpha_n} \right\|_{2}^{2} 
			= \int_{\mathbb{R}^2 \setminus B_{\sqrt{\alpha_n}}(\mathbf{0})} \hspace{-0.3cm} d\mathbf{x}\,\left| \mathbf{A}_{\alpha_n} \eta_{\alpha_n} \psi_{0} \right|^{2} +\! \int_{B_{\sqrt{\alpha_n}}(\mathbf{0})}\hspace{-0.3cm} d\mathbf{x}\,\left| \mathbf{A}_{\alpha_n} \eta_{\alpha_n} \psi_{0} \right|^{2} .
	\end{equation*}
By elementary estimates we get
	\begin{equation*}
		\int_{\mathbb{R}^2 \setminus B_{\sqrt{\alpha_n}}(\mathbf{0})}\hspace{-0.3cm} d\mathbf{x}\,\left| \mathbf{A}_{\alpha_n} \eta_{\alpha_n} \psi_{0} \right|^{2} \leqslant \alpha_n \left\|\psi_{0} \right\|_{2}^{2}\,.
	\end{equation*}
On the other side, keeping in mind that $\psi_0 \!\in\! H^1(\mathbb{R}^2)$, we use a sharp result on Sobolev embeddings \cite{ET99} and dominated convergence to infer
	\begin{align*}
		& \int_{B_{\sqrt{\alpha_n}}(\mathbf{0})}\hspace{-0.3cm} d\mathbf{x}\,\left| \mathbf{A}_{\alpha_n} \eta_{\alpha_n} \psi_{0} \right|^{2} 
			= \int_{B_{\sqrt{\alpha_n}}(\mathbf{0})}\hspace{-0.3cm} d\mathbf{x}\;{\alpha_n^2 \over |\mathbf{x}|^2} \left({|\mathbf{x}| \over \sqrt{\alpha_n}}\right)^{2\alpha_n} \left|\psi_{0} \right|^{2} \\
		& \leqslant \alpha_n^{2 -\alpha_n} \esssup_{\mathbf{x} \in B_{\sqrt{\alpha_n}}(\mathbf{0})}\! \left(\, |\mathbf{x}|^{2\alpha_n} \big(1 + |\log|\mathbf{x}|\,|\,\big)^2 \right) \int_{B_{\sqrt{\alpha_n}}(\mathbf{0})}\hspace{-0.3cm} d\mathbf{x}\; {\left|\psi_{0} \right|^{2} \over |\mathbf{x}|^2 \big(1 + |\log|\mathbf{x}|\,|\big)^2} \\
		& \leqslant e^{ -2 - \alpha_n \log \alpha_n + 2\alpha_{n}} \int_{B_{1}(\mathbf{0})}\hspace{-0.2cm} d\mathbf{x}\; {\mathbf{1}_{B_{\sqrt{\alpha_n}}(\mathbf{0})}\; \left|\psi_{0} \right|^{2} \over |\mathbf{x}|^2 \big(1 + |\log|\mathbf{x}|\,|\,\big)^2} \;\xrightarrow{n \to +\infty}\; 0\,.
	\end{align*}
The above arguments show that
	\begin{equation}\label{eq:An0}
		\left\| \mathbf{A}_{\alpha_n} \psi_{\alpha_n} \right\|_{2}^{2} \;\xrightarrow{n \to +\infty}\; 0\,.
	\end{equation}
Next, let us examine the behavior in $H^1(\mathbb{R}^2)$ of the sequence $\{\psi_{\alpha_n}\}_{n \,\in\, \mathbb{N}}$, taking into account that we already established strong convergence $\psi_{\alpha_n} \!\to\! \psi_0$ in $L^2(\mathbb{R}^2)$. By triangular inequality, we get
	\begin{equation*}
		\left\| \nabla \psi_{\alpha_n}\!- \nabla \psi_{0}  \right\|_{2}
		\leqslant \left\|(\eta_{\alpha_n} \!- 1) \nabla \psi_0 \right\|_{2} + \left\| (\nabla \eta_{\alpha_n})\,\psi_0 \right\|_{2} .
	\end{equation*}
Recalling once more that $\psi_0 \!\in\! H^1(\mathbb{R}^2)$, by dominated convergence we obtain
	\begin{equation*}
		\left\|(\eta_{\alpha_n} \!- 1) \nabla \psi_0\right\|_2^{2} \,= \int_{\mathbb{R}^2}\! d\mathbf{x}\;\left|\eta_{\alpha_n} - 1 \right|^2\, \left|\nabla\psi_{0}\right|^2 \;\xrightarrow{n \to +\infty}\; 0\,.
	\end{equation*}
On the other hand, from \eqref{eq:eta01} we deduce
	\begin{equation*}
		\left|\nabla \eta_{\alpha_n}(\mathbf{x}) \right| = \left\{\, \begin{array}{ll}
			\displaystyle{\sqrt{\alpha_n}\left(\,|\mathbf{x}|/\sqrt{\alpha_n}\,\right)^{\alpha_n-1} = \left|\mathbf{A}_{\alpha_n}(\mathbf{x})\right| \eta_{\alpha_n}(\mathbf{x})} & \quad \mbox{for\, $\mathbf{x} \!\in\! B_{\sqrt{\alpha_n}}(\mathbf{0})$}\,, \vspace{0.1cm}\\
			\displaystyle{0} & \quad \mbox{for\, $\mathbf{x} \!\in\! \mathbb{R}^2 \!\setminus\! B_{2\sqrt{\alpha_n}}(\mathbf{0})$}\,;
		\end{array}\right.
	\end{equation*}
the downward concavity of $\eta_{\alpha_n}$ further ensures
	\begin{equation*}
		\left|\nabla \eta_{\alpha_n}(\mathbf{x}) \right| \leqslant \sqrt{\alpha_n}
	 	\qquad \mbox{for\, $\mathbf{x} \in B_{2\sqrt{\alpha_n}}(\mathbf{0}) \!\setminus\! B_{\sqrt{\alpha_n}}(\mathbf{0})$}\,.
	\end{equation*}
The above relations, together with \eqref{eq:An0}, entail
	\begin{align*}
		\left\| (\nabla \eta_{\alpha_n})\,\psi_0 \right\|_{2}^2 
		& \leqslant \!\int_{B_{\sqrt{\alpha_n}}(\mathbf{0})}\hspace{-0.4cm} d\mathbf{x}\; \left|\mathbf{A}_{\alpha_n}(\mathbf{x})\right|^2 \left|\eta_{\alpha_n}(\mathbf{x}) \psi_0 \right|^2  
			+ \alpha_n\! \int_{B_{2\sqrt{\alpha_n}}(\mathbf{0}) \setminus B_{\sqrt{\alpha_n}}(\mathbf{0})}\hspace{-0.5cm} d\mathbf{x}\, \left|\psi_0 \right|^2 \\
		& \hspace{3.7cm} \leqslant \left\| \mathbf{A}_{\alpha_n} \psi_{\alpha_n} \right\|_{2}^{2} 
			+ \alpha_n \left\|\psi_{0} \right\|_{2}^{2} \;\xrightarrow{n \to +\infty}\; 0\,.
	\end{align*}
Summing up, we have $\left\| \nabla \psi_{\alpha_n}\!- \nabla \psi_{0}  \right\|_{2} \!\to\! 0$ for $n \!\to\! +\infty$, which entails strong convergence $\psi_{\alpha_n}\! \to \psi_{0}$ in $H^1(\mathbb{R}^2)$. In particular, we have that $\{\psi_{\alpha_n}\}_{n \,\in\, \mathbb{N}}$ is a uniformly bounded sequence in $H^1(\mathbb{R}^2)$.\\
Returning to \eqref{eq:teles} and recalling that $\mathbf{S} \!\in\!L^{\infty}(\mathbb{R}^2)$, on account of the results derived above we finally obtain
	\begin{align*}
		& \left|\,Q_{\alpha_n,S}^{(F)}[\psi_{\alpha_n}] - Q_{0,S}^{(F)}[\psi_0] \,\right| \\
		& \leqslant C \left[ \left\| \mathbf{A}_{\alpha_n} \psi_{\alpha_n} \right\|_{2}^{2}
			+ \left\|\psi_{\alpha_{n}} \right\|_{H^1} \left\| \mathbf{A}_{\alpha_n} \psi_{\alpha_{n}} \right\|_{2}
			+ \left(\, \left\|\psi_{\alpha_{n}} \right\|_{H^1}\! + \left\|\psi_0 \right\|_{H^1} \right) \left\| \psi_{\alpha_{n}} \!- \psi_{0} \right\|_{H^1} \right] \\
		& \hspace{9.8cm} \xrightarrow{n \to +\infty}\; 0\,,		
	\end{align*}
which proves \eqref{eq:proofx}, whence the thesis \eqref{eq:upperb}.
\end{proof}

\begin{acknowledgement}
I wish to thank Michele Correggi for stimulating conversations and valuable comments on the content of this work.

This work has been supported by: the European Research Council (ERC) under the European Union’s Horizon 2020 research and innovation programme (ERC CoG UniCoSM, grant agreement n. 724939); INdAM-GNFM Progetto Giovani 2020 {\sl Emergent Features in Quantum Bosonic Theories and Semiclassical Analysis}; Istituto Nazionale di Alta Matematica ``F. Severi'', through the Intensive Period ``INdAM Quantum Meetings (IQM22)''.
\end{acknowledgement}

%%%%%%%%%%%%%%%%%%%%%%%% referenc.tex %%%%%%%%%%%%%%%%%%%%%%%%%%%%%%
% sample references
% %
% Use this file as a template for your own input.
%
%%%%%%%%%%%%%%%%%%%%%%%% Springer-Verlag %%%%%%%%%%%%%%%%%%%%%%%%%%
%
% BibTeX users please use
% \bibliographystyle{}
% \bibliography{}
%
% \biblstarthook{}

\end{document}